\documentclass[a4paper,fleqn,usenatbib]{mnras}

\usepackage{newtxtext,newtxmath}
\usepackage[T1]{fontenc}
\usepackage{ae,aecompl}

\usepackage{graphicx}	% Including figure files
\usepackage{amsmath}	% Advanced maths commands
\usepackage{amssymb}	% Extra maths symbols
\usepackage{siunitx}
\usepackage{tabularx}   % MJ: to spread out table over whole page

\title[Observing substructure in discs around MYSOs]{Observing substructure in circumstellar discs around massive young stellar objects}

\author[M. R. Jankovic et al.]
{\parbox{\textwidth}{M.~R.~Jankovic$^{1}$\thanks{E-mail: \texttt{m.jankovic16@imperial.ac.uk}},
T.~J.~Haworth$^{1}$,
J.~D.~Ilee$^{2,3}$,
D.~H.~Forgan$^{4}$,
C.~J.~Cyganowski$^{4}$,\\
C.~Walsh$^{3}$,
C.~L.~Brogan$^{5}$,
T.~R.~Hunter$^{5}$
and S.~Mohanty$^{1}$%, (final authors/order TBD)
}\vspace{0.4cm}\\
\parbox{\textwidth}{$^{1}$Astrophysics Group, Imperial College London, Blackett Laboratory, Prince
Consort Road, London SW7 2AZ, UK\\
$^{2}$Institute of Astronomy, Madingley Road, Cambridge CB3 0HA, UK \\
$^{3}$School of Physics and Astronomy, University of Leeds, Leeds LS2 9JT, UK \\
$^{4}$SUPA, School of Physics \& Astronomy, University of St Andrews, North Haugh, St Andrews, Scotland, KY16 9SS, UK \\
$^{5}$NRAO, 520 Edgemont Rd, Charlottesville, VA 22903, USA }}

\date{Accepted XXX. Received YYY; in original form ZZZ}

\pubyear{2017}

\begin{document}
\label{firstpage}
\pagerange{\pageref{firstpage}--\pageref{lastpage}}
\maketitle

\begin{abstract}

Simulations of massive star formation predict the formation of discs with significant substructure, such as spiral arms and clumps due to fragmentation.  Here we present a semi-analytic framework for producing synthetic observations of discs with substructure, in order to determine their observability in interferometric observations.  Unlike post-processing of hydrodynamical models, the speed inherent to our approach permits us to explore a large parameter space of star and disc parameters, and thus constrain properties for real observations.   
We compute synthetic dust continuum and molecular line observations probing different disc masses, distances, inclinations, thermal structures, dust distributions, and number and orientation of spirals and fragments. With appropriate spatial and kinematic filtering applied, our models predict that ALMA observations of massive YSOs at $<5$\,kpc distances should detect spirals in both gas and dust in strongly self-gravitating discs (i.e. discs with up to two spiral arms and strong kinematic perturbations). Detecting spirals will be possible in discs of arbitrary inclination, either by directly spatially resolving them for more face-on discs (inclinations up to $\sim$\,50 degrees), or through a kinematic signature otherwise. Clumps resulting from disc fragmentation should be detectable in the continuum, if the clump is sufficiently hotter than the surrounding disc material.

\end{abstract}


%%%%%%%%%%%%%%%%%%%%%%%%%%%%%%%%%%%%%%%%%%%%%%%%%%

%%%%%%%%%%%%%%%%% BODY OF PAPER %%%%%%%%%%%%%%%%%%

\section{Introduction}

Circumstellar discs have long been detected around low to intermediate mass ($M_*\lesssim8$\,M$_\odot$) young stars.  These discs are known to accrete material on to the star, and are of increasing interest as the sites of planet formation \citep[e.g.][]{2011ARA&A..49...67W, 2015arXiv150906382A, 2016JGRE..121.1962M, 2016PASA...33...13C}. In the higher mass regime ($M_*\gtrsim8$\,M$_\odot$) the very short pre-main-sequence lifetime ($<10$\,Myr), coupled with the embedded, distant ($\geq$kpc) and relatively scarce nature of such stars, makes the detection of discs more difficult. Nevertheless, there is growing observational evidence for the presence of discs around massive YSOs \citep{2014ApJ...788..187H, 2015ApJ...813L..19J, 2015MNRAS.447.1826Z,  2016ApJ...823..125C, 2016MNRAS.462.4386I}. 

\smallskip

Investigating the nature and properties of discs around massive stars is important for a number of reasons. Firstly, as a star gains mass, radiative and kinematic (i.e. stellar wind) feedback drives ambient material away and could effectively shut off accretion and hence limit the potential stellar mass \citep{1974A&A....37..149K, 1989ApJ...345..464N, 2004MNRAS.349..678E}. The fact that we observe very massive stars \citep[e.g.][]{massey_2001} requires some mechanism that permits the star to continue to accrete. Numerical simulations have demonstrated that feedback can be overcome by accretion through a disc \citep[e.g.][]{2010ApJ...722.1556K, 2011ApJ...732...20K, 2016ApJ...823...28K, 2016MNRAS.463.2553R, 2017MNRAS.471.4111H}, with the feedback energy escaping through the poles.  Therefore, circumstellar discs are pivotal in allowing massive stars to grow, and so characterising them observationally is key.

\smallskip

Secondly, if gravitationally unstable, a circumstellar disc can produce substructure which can affect the accretion on to the star, and even fragment into gravitationally bound objects (discussed further below). A Keplerian disc will be gravitationally unstable when the \cite{1964ApJ...139.1217T} $Q$ parameter satisfies
\begin{equation}
	Q \equiv \frac{c_{\rm s}\Omega}{\pi G \Sigma} <1
	\label{equn:ToomreQ}
\end{equation}
where $c_{\rm s}$, $\Omega$ and $\Sigma$ are the sound speed, angular frequency, $(GM_*/R^3)^{1/2}$ ,and surface density at a distance $R$ from a star of mass $M_*$. If we assume that the disc mass scales linearly with the stellar mass, and the temperature in the deeper layers of the outer disc is roughly constant at $\sim$10\,K (i.e. is independent of stellar mass) we would hence expect from equation (\ref{equn:ToomreQ}) that $Q\propto M_*^{-1/2}$.  Under this simple argument discs around more massive YSOs are more susceptible to becoming gravitationally unstable \citep[though instability in discs around lower mass stars certainly can arise;][]{2016Sci...353.1519P}. Gravitational instability is therefore something that is important to consider in the study of discs around massive YSOs. 

\smallskip

Numerical models of gravitationally unstable discs predict spiral overdensities and fragmentation of the disc to form clumps \citep{2017MNRAS.470.2517H,2006MNRAS.373.1563K}. This instability of the disc may drive up the accretion rate compared to regular viscous evolution \citep{1994ApJ...436..335L}.  Such an effect may provide an explanation for recent outbursting behaviour detected toward several massive young stellar objects \citep{hunter_2017, caratti_2017}. Furthermore, the resulting clumps may be the seeds of additional gravitationally bound objects such as stars or planets \citep{2018MNRAS.474.5036F}. If bound to the massive primary, future interaction (i.e. accretion) of such secondary objects can have consequences for the later stages of massive stellar evolution \cite[for a review see][]{2017PASA...34....1D}.

\smallskip

Observationally, however, it is extremely difficult to detect the immediate circumstellar environments of massive young stars due to their embedded nature, their relative scarcity, and their correspondingly large distances.  Infrared    (IR) interferometry    and  high-resolution near-infrared spectroscopy   have revealed discs on  scales of less than  $1000$\,AU around  massive YSOs \citep{kraus_2010,   boley_2013, bik_2004, wheelwright_2010, ilee_2013, 2017A&A...604A..78R} but  these techniques  are limited to tracing  only the inner regions of discs, and provide little information on the bulk of the circumstellar environment. %are limited  to scales of tens of AU around relatively    evolved,     IR-bright    objects.  
Longer-wavelength interferometric observations allow  access to the circum(proto)stellar environments of  less evolved,  more embedded  massive YSOs, but often probe larger spatial scales.  In many cases velocity gradients detected in millimetre and centimetre-wavelength molecular line observations trace large-scale  (1000s to $\gtrsim$10,000\,AU),  massive ($\mathrm{M}_{\mathrm{toroid}} \ge \mathrm{M}_{*}$), non-equilibrium   rotating   structures known as `toroids' \citep[e.g.][and   references therein]{cesaroni_2005_apss,cesaroni_2006,cesaroni_2007,beuther_2008,beltran_2011,cesaroni_2011,johnston_2014}.  Nevertheless, high angular resolution observations are beginning to provide tantalizing evidence for Keplerian discs around massive protostars \citep[e.g.][]{2014ApJ...788..187H, 2015ApJ...813L..19J, 2015MNRAS.447.1826Z,  2016ApJ...823..125C, 2016MNRAS.462.4386I}.  

\smallskip

It is therefore imperative to develop a robust framework within which we are able to extrapolate a robust physical interpretation from such observations.  Synthetic observations, where the appearance of a model to an observer is computed (ideally including the instrumentation response), are particularly valuable in this regard \citep[for a review see][]{2018NewAR..82....1H}.  To date, some synthetic observations have been computed from dynamical simulations of massive star formation/disc evolution, and are generally very optimistic about the detection of substructure in discs \citep[e.g.][]{2007ApJ...656..959K, 2017MNRAS.471.4111H, 2018MNRAS.473.3615M}. However, these studies typically place objects at nearby distances ($\leq$1\,kpc), do not always account for the full instrumentational (e.g. interferometric) effects, and are typically concerned with predicting the dust continuum emission from these objects, rather than observations of molecular lines that can trace the gas emission and kinematics.  In addition, approaches involving full two or three dimensional (radiation-)hydrodynamics are extremely computationally expensive, and thus only permit a small region of parameter space to be explored.    The main caveat of existing synthetic observations of discs around massive YSOs is that they are only able to focus on a small number of scenarios, owing to the cost of the dynamical simulations. Given the huge range of star-disc parameters, such as the stellar and disc mass, thermal and chemical structure, number of spirals/clumps, and so on, a less expensive means of producing models from which to generate synthetic observations is extremely valuable.

\smallskip

In this paper, we combine semi-analytic models of self-gravitating discs with radiative transfer models to provide forward modelling predictions for upcoming Atacama Large Millimetre Array (ALMA) observations, in order to gauge the impact of different physical conditions on the observability of massive discs and their substructure.  We note that the framework we present can be easily adapted to perform retrieval modelling of observations.  This paper is organized as follows. In Section \ref{sec:methodology} we describe how we construct the semi-analytic models of self-gravitating discs, our radiative transfer calculations and how we account for the instrumentational interferometric effects. In Section \ref{sec:results} we present an overview of the challenges that arise in the spatial and kinematic detection of substructure in our fiducial disc model, and how we use fast and efficient filtering techniques in order to enhance this substructure.  In Section \ref{sec:discussion} we explore and discuss critical effects on substructure observability, and make recommendations for future observational campaigns.  Finally, in Section \ref{sec:conclusions} we summarize our main conclusions.

\section{Methodology}
\label{sec:methodology}

We begin by describing how we construct our discs and process them to produce synthetic ALMA observations. There are three phases to this process. First we set up the semi-analytic disc structure (as detailed in section \ref{sec:discStructure}). Next, we compute the properties of the dust and the abundance of the molecule, and a radiative transfer calculation is used to compute the flux received by an observer (section \ref{sec:RT}).  Finally we account for instrumentational effects inherent to interferometry and produce simulated observations of the models in both continuum and molecular line emission (section \ref{sec:casa}).

\begin{figure*}
    \includegraphics[width=0.32\textwidth]{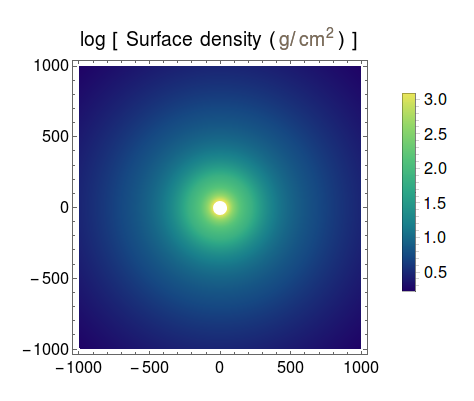}
    \includegraphics[width=0.32\textwidth]{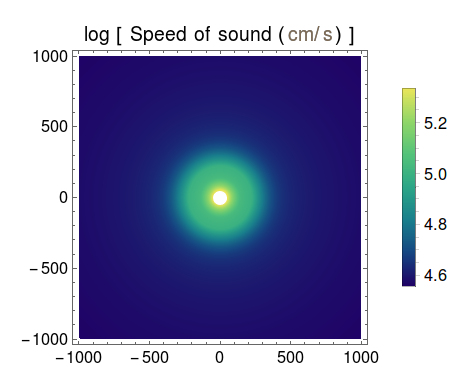}
    \includegraphics[width=0.32\textwidth]{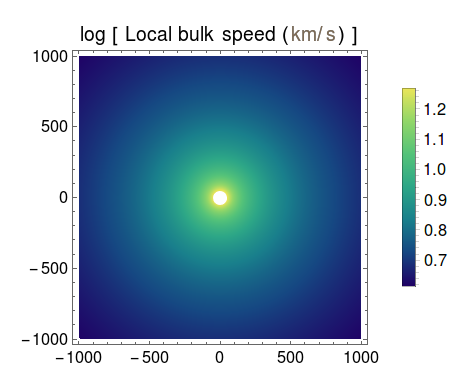}
    \includegraphics[width=0.32\textwidth]{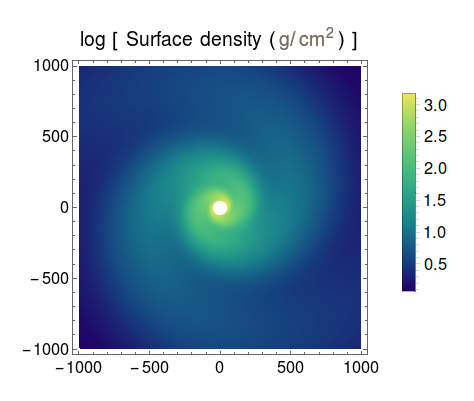}
    \includegraphics[width=0.32\textwidth]{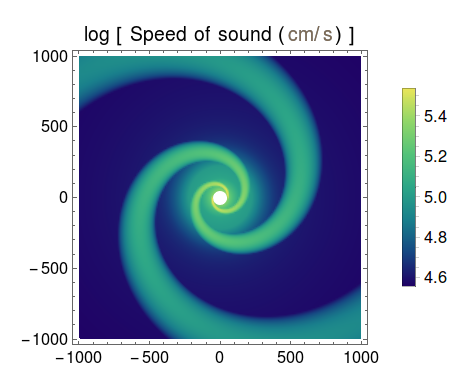}
    \includegraphics[width=0.32\textwidth]{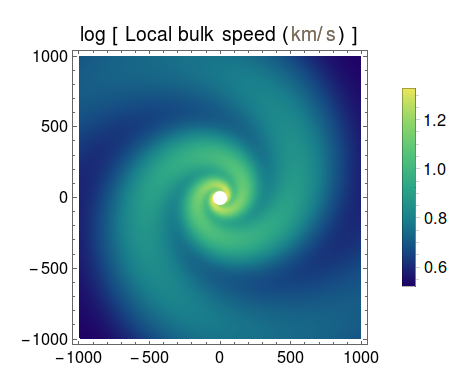}
    \includegraphics[width=0.32\textwidth]{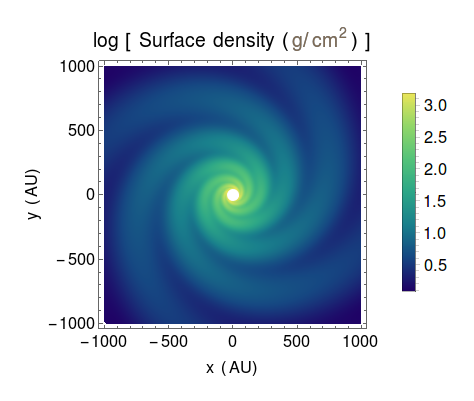}
    \includegraphics[width=0.32\textwidth]{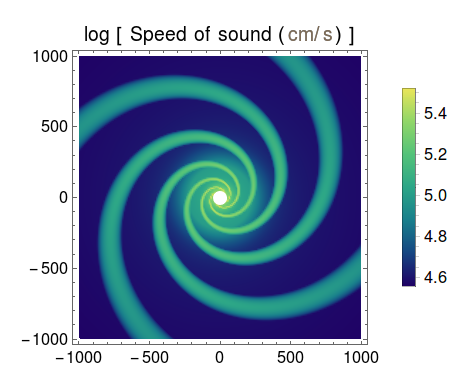}
    \includegraphics[width=0.32\textwidth]{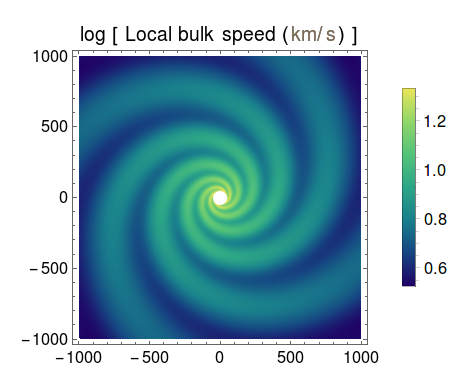}
    \caption{Self-gravitating discs constructed using the two-dimensional semi-analytic model: axisymmetric (top row), 2 spiral arms (middle row), 4 spiral arms (bottom row). The panels show the surface density, sound speed and bulk velocity from left to right. }
    \label{discplot}
\end{figure*}

\subsection{Disc construction }
\label{sec:discStructure}
The circumstellar matter surrounding our massive YSOs is constructed using a simple semi-analytic approach, following \cite{2009MNRAS.396.1066C},  \cite{2009MNRAS.396.2228R}, and \cite{2016MNRAS.463..957F}.  In this framework, we set up a radial profile of an axisymmetric quasi-steady self-gravitating disc. Upon this we then impose a spiral structure as a perturbation (following the procedure of \citealt{2016MNRAS.458..306H}), and in some models include the presence of a fragment within the disc.  

\subsubsection{Underlying axisymmetric radial structure of the disc }
From the inner ($r_{\rm in}=50\,$AU) to the outer ($r_{\rm out}=1000$\,AU) boundary of the disc, at each radius we iteratively solve a system of equations that determines the disc structure. Solutions are determined by the stellar mass $M_*$ and the accretion rate $\dot{M}$. We fix the stellar mass to $M_* = 20$\,M$_{\odot}$ in this paper, but the accretion rate $\dot{M}$ is varied to produce a set of corresponding quasi-steady state disc solutions (i.e. $\dot{M}$ is constant at all radii).

\smallskip

Starting with an initial estimate for the surface density $\Sigma$, we calculate the angular velocity $\Omega$ at radius $r$
\begin{equation}
    \Omega = \sqrt{\frac{G (M_* + \int_{r_{\rm in}}^{r} 2\pi r \Sigma dr)}{r^3}}.
\end{equation}

Numerical simulations predict that self-gravitating discs settle into a marginally stable state where the Toomre $Q$ parameter \citep{1964ApJ...139.1217T} is roughly constant throughout the disc and close to $\sim1.7$ \citep{durisen_2007}. %\citep[e.g.][]{LodatoRice2004, LodatoRice2005}. %DHF: If Q \leq 2, and the disc isn't fragmenting, you will see spiral structure.  That is something we can say with reasonable confidence.  The difference between Q=2 and Q=1 is merely a matter of degree.  Q=2 is the minimum criterion for nonaxisymmetric instability (spirals).  Q=1 guarantees instability even for axisymmetric (ring) perturbations.  In numerical simulations, most discs settle into a marginally stable regime at Q~1.7 (Durisen et al 2007, PPV).
Thus, following \citet{2009MNRAS.396.2228R}, \citet{2016MNRAS.463..957F} and \citet{2016MNRAS.458..306H}, we impose a constant value of the Toomre parameter
\begin{equation}
    Q = \frac{c_{\rm{s}} \kappa_{\rm{epi}}}{\pi G \Sigma} = 2,
\end{equation}
where $\kappa_{\rm{epi}}$ is the epicyclic frequency, and $c_{\rm{s}}$ is the local sound speed. For Keplerian discs $\kappa_{\rm{epi}} = \Omega$, and we assume the same here. From the last expression we then find the sound speed $c_{\rm{s}}$ and from it the disc scale height $H$, where
\begin{equation}
    H = \frac{c_{\rm{s}}}{\Omega},
\end{equation}
and mid-plane density
\begin{equation}
    \rho_0 = \frac{\Sigma}{2 H}.
\end{equation}

Given the sound speed $c_{\rm{s}}$ and mid-plane density $\rho_0$, we use the equations of state table \citep{2007A&A...475...37S, 1975ApJ...199..619B} to find the temperature. We use the temperature $T$ and the density $\rho_0$ to find the mass-mean opacity $\kappa$ \citep{2007A&A...475...37S}, and the optical depth from the surface of the disc to the disc midplane, at the given radius,
\begin{equation}\label{eq:tau}
    \tau = \frac{1}{2} \kappa \Sigma.
\end{equation}
The mass-mean opacity $\kappa$ is obtained by averaging the Rosseland-mean opacity from \cite{1994ApJ...427..987B} over a spherically symmetric polytropic pseudo-cloud. This takes into account effects of the surrounding warmer/colder matter.

\smallskip

Following \cite{2009MNRAS.396.2228R} and \cite{2016MNRAS.463..957F}, we assume that angular momentum transport can be approximated as local and pseudo-viscous. The accretion rate $\dot{M}$ is then given by
\begin{equation}\label{eq:Mdot}
    \dot{M} = 3 \pi \nu \Sigma = \frac{3 \pi \alpha c_{\rm{s}}^2 \Sigma}{\Omega}
\end{equation}
where $\nu$ is the viscosity and $\alpha$ is the pseudo-viscous parameter \citep{1973A&A....24..337S}. The two are related by $\nu = \alpha c_{\rm s} H$. In quasi-steady state $\dot{M}$ is constant at all radii.

\smallskip

The unknown pseudo-viscous parameter $\alpha$ is found by assuming that the strength of the angular momentum transport is set by the cooling rate. The cooling rate is given by \citep{2009MNRAS.396.2228R}
\begin{equation}\label{eq:Lambda}
    \Lambda = \frac{16 \sigma_{\rm{SB}}}{3}\frac{(T^4-T_{\rm{irr}}^4)}{\tau + \tau^{-1}},
\end{equation}
where $\sigma_{\rm SB}$ is the Stefan-Boltzmann constant and $T_{\rm{irr}}$ is the temperature in the disc due to irradiation from the star, and we assume that gas and dust temperatures are the same. We find that using the standard Stefan-Boltzmann relationship between the stellar luminosity \citep[$L_* \sim 7.5 \times 10^4$\,L$_\odot$ for the stellar mass $M_* = 20$\,M$_\odot$,][]{hosokawa_2009}, distance from the star, and the equilibrium irradiation temperature, $T_{\rm irr}$, yields unrealistically high disc temperatures compared to results from hydrodynamical simulations with detailed radiative transfer (T. Douglas, T. Harries, priv. comm.). This is possibly due to the stellar irradiation being heavily reprocessed in the outer disc. Thus, we follow \citet{2009MNRAS.396.2228R} and assume a constant $T_{\rm irr}=10$\,K throughout the disc, bringing our results into agreement with the detailed simulations. The stellar irradiation should become much more important in the innermost disc ($r \lesssim 50$\,AU), where the line-of-sight optical depth from the star to the disc midplane is much smaller than in the outer disc. However, given the observational parameters that we probe in this work, the innermost disc is not resolved in our synthetic observations. Thus, we do not model the highly irradiated inner disc and artificially set the inner boundary of the disc to $r_{\rm in}=50$\,AU. Finally, if the irradiation dominates over the pseudo-viscous heating ($T_{\rm{irr}} > T$), we set $T = T_{\rm{irr}}$. For our fiducial accretion rate $\dot{M}=10^{-3}$\,M$_\odot$\,yr$^{-1}$, $T > T_{\rm{irr}}$ throughout the disc.

\smallskip

The cooling rate, given by equation (\ref{eq:Lambda}), is balanced against (equal to) the viscous dissipation rate
\begin{equation}
    D = \frac{9}{4} \nu \Sigma \Omega^2 = \frac{9}{4} \alpha c_{\rm{s}}^2 \Sigma \Omega,
\end{equation}
which we use to find the pseudo-viscous parameter $\alpha$. Next, we can find the accretion rate $\dot{M}$ from equation (\ref{eq:Mdot}) and compare it to the value for which we want to solve the structure of the steady-state disc, based on which we iteratively improve the value of the surface density $\Sigma$.

\smallskip

\subsubsection{Spiral substructure} 
\label{sec:spiralsetup}

The radial model described above gives us an axisymmetric disc. We impose a spiral structure on that disc by adding perturbations to it. The surface density is perturbed as in \cite{2016MNRAS.458..306H}, using results of \cite{2009MNRAS.393.1157C}
\begin{equation} \label{eq:sigma_perturbation}
    \Sigma' = \Sigma - \alpha^{1/2} \Sigma \cos( m (\Theta(r) - \phi + \theta) )
\end{equation}
where $\phi=\tan (y/x)$, $m$ is the number of spiral arms, and $\Theta(r)$ determines how tightly the spirals are wrapped. We set $\Theta(r) = \frac{1}{b} \log(\frac{r}{a})$, $a = 13.5$ and $b = 0.38$, the same as \cite{2016MNRAS.458..306H}. The angle $\theta$ is the phase offset of the spiral, which we introduce in order to investigate how the angular orientation of the spirals with respect to the observer, at the time of the observation, might influence the ability to resolve the spiral arms.

\smallskip

In this work we also vary the number of spiral arms $m$. Numerical simulations show that lower-mass ($M_{\rm d}/M_*<0.1$) self-gravitating discs develop a large number of spiral arms \citep[$m > 10$;][]{2009MNRAS.393.1157C}, while higher-mass discs ($M_{\rm d}/M_*>0.25$) develop fewer \citep[$m \sim 2$;][]{LodatoRice2004, LodatoRice2005, forgan_2011}.  For our fiducial disc accretion rate $\dot{M}=10^{-3}$\,M$_\odot$\,yr$^{-1}$ the semi-analytic 1D disc model yields a disc-to-star mass ratio of $\sim 0.3$, placing it in the regime of massive discs with few spiral arms. We thus set $m=2$ in our fiducial disc model, and also run a model with $m=4$ spiral arms to probe how the spatial separation between the spirals affects the possibility of their detection.

\smallskip

Finally, if $\alpha$ obtained from the 1D model is above a certain value $\alpha_{\rm{sat}} = 0.1$, at which gravitational torque saturates \citep{2001ApJ...553..174G, 2005MNRAS.364L..56R, 2017ApJ...847...43D}, here we set it to $\alpha = \alpha_{\rm{sat}}$.
 
\smallskip

Here we are also interested in kinematic effects, so we also perturb the angular velocity of matter in the disc. If we assume that the radial velocity of matter is negligible compared to the angular component, we can use the angular projection of the continuity equation to show that the velocity perturbation is proportional to the density perturbation. We neglect the radial velocity $\dot{M}/(2\pi r \Sigma)$ in our models, although we note that it can be up to a few tenths of the Keplerian velocity in the outermost disc. The sign of the angular velocity perturbation then depends on the velocity of the spiral pattern compared to the local Keplerian velocity - it is positive if the spiral pattern is rotating faster than the unperturbed velocity, and negative in the opposite case. Therefore, we perturb the angular velocity via
\begin{equation}
    \Omega' = \Omega - \eta \Omega \cos( m (\Theta(r) - \phi + \theta) )
    \label{equn:spiralPerturb}
\end{equation}
where $\eta$ is a free parameter that controls the magnitude of the perturbation. The $\eta$ parameter is determined by the rotation velocity of the spiral pattern which is roughly proportional to the disc-to-star mass ratio \citep{2009MNRAS.393.1157C, forgan_2011}. We set $\eta=0.2$ in our fiducial disc model, appropriate given the disc-to-star mass ratio in the model, but also vary $\eta$ to probe the observational effect of weaker perturbations.

\smallskip

Furthermore, we set the perturbed sound speed to the value of the local fluid velocity relative to the unperturbed speed $c_{\rm{s}}' = |\Omega' - \Omega| r$ (with a lower limit of the unperturbed sound speed, $c_{\rm{s}}$). This accounts for the higher temperatures expected in the spirals due to heating from shocks \citep{2009MNRAS.393.1157C}.

\smallskip

An illustration of our semi-analytic discs is given in Figure \ref{discplot}, which compares the surface density, sound speed and local bulk speed of axisymmetric, 2-arm spiral and 4-arm spiral discs.

\subsubsection{Prescription for fragments}
\label{sec:fragment_method}

In addition to discs with spiral structure, we also explore the observational characteristics of discs undergoing fragmentation. To that end, we insert clumps within the disc to represent the presence of a fragment within our disc models.  The clump is modelled as an isothermal Gaussian within a sphere of radius $R_{\rm C}$
\begin{equation}
    \rho = \rho_{\rm C} \exp \left( -\frac{ |\vec{r} - \vec{r_{\rm C}}|^2 }{ 2 R_{\rm C}^2 } \right).
\end{equation}
where $\vec{r}$ is the vector relative to the clump centre. We assign the clump temperature $T_{\rm C}$, central density $\rho_{\rm C}$ and radius $R_{\rm C}$ guided by the analysis of fragments in smoothed particle hydrodynamics simulations by \cite{2017MNRAS.470.2517H}. The center of the clump $\vec{r_{\rm C}}$ in our models is always set so that the clump is located within a spiral arm (as is often the case in simulations of fragmenting discs, see e.g., \citealt{boley_2009}).

\subsection{Radiative transfer modelling}
\label{sec:RT}

We produce synthetic images and molecular line datacubes from the above disc models using the \textsc{torus} radiative transfer and hydrodynamics code \citep[e.g.][]{2000MNRAS.315..722H, 2004MNRAS.351.1134K, 2012MNRAS.420..562H}, with the molecular-line transfer ray-tracing scheme presented by \cite{2010MNRAS.407..986R}. We now summarise the relevant details of these calculations. 

\subsubsection{Constructing a 3D disc }
\label{sec:3d_disc}

We map the 2D models, described in section \ref{sec:discStructure}, onto the mid-plane of a \textsc{torus} grid using bilinear interpolation. We then construct the vertical structure of the disc assuming hydrostatic equilibrium. We give ourselves the freedom to explore the effect of a disc that is not isothermal using a parametric temperature profile of the form:
\begin{equation}
    T(R, z) = \left\{ 
    \begin{array}{cc}
         T_{\rm{mid}} + (T_{\rm{atm}} - T_{\rm{mid}})\left[\sin\left(\frac{\pi z}{2z_{\rm{q}}}\right)\right] & z < z_{\rm{q}}\\
          T_{\rm{atm}} & z\geq z_{\rm{q}}
    \end{array}
     \right.
\end{equation}
following \cite{2014ApJ...788...59W} where we set $T_{\rm{atm}}=f_{\rm atm} T_{\rm{mid}}$ and $z_{\rm{q}}$ is four times the disc scale height. This allows us to compare an isothermal disc ($f_{\rm atm}=1$) with increasing amounts of vertical heating ($f_{\rm atm}>1$) that could have important observational impact (for example sublimating dust, or destroying or exciting molecules).

\subsubsection{Dust continuum emission}
\label{sec:dust}
Dust is included in our radiative transfer models, contributing as a continuum emission source and an opacity source to the line emission. We assume a single distribution of grains at all radii that is well mixed with the gas with a dust-to-gas mass ratio of 10$^{-2}$, though we also follow \cite{2009ApJ...702..567V} and sublimate dust in regions where the temperature is above 1500\,K. The dust distribution is a \cite{1977ApJ...217..425M} power law ($n(a)\propto a^{-q}$) specified in terms of the minimum and maximum size and a power law for the grain size distribution. For most of the models in this paper we use $a_{\rm min}=1$\,nm, $a_{\rm max}=1$\,cm and  $q=3.3$, though we probe the impact of grain growth in section \ref{sec:optical_depth}. 

\subsubsection{Gas emission and molecular abundances}
\label{sec:molecule_prescription}

We have chosen to focus our observational predictions for the gas on the CH$_{3}$CN $J=13$--12 ladder of transitions. Emission from CH$_{3}$CN is widely detected in the vicinity of massive protostars \citep[see e.g.][]{hunter_2014, 2016MNRAS.462.4386I, cesaroni_2017, beuther_2017}.  The ladder is an incredibly useful diagnostic of the physical conditions of the gas, due to the large number of K transitions that span a range of excitation energies, and the fact that they are sufficiently closely-spaced to enable simultaneous observation. Within the $J=13$--12 ladder, transitions range from the K=0 transition at 239.138\,GHz (E$_{\rm up} = 80$\,K) to the K=12 transition at 238.478\,GHz (E$_{\rm up} = 1106$\,K). %A given transition of a molecule will emit more or less strongly depending on the local temperature and density of the emitting material. 
We have chosen to concentrate on the K=3 transition (%239.0965
239.096\,GHz, E$_{\rm up} = 145$\,K) as this line traces the entire radial extent of the disc in our models.

\smallskip

We assume local thermodynamic equilibrium (LTE), which significantly reduces the computational cost of the models compared to non-LTE line transfer. The CH$_3$CN abundance is canonically assumed to be $10^{-7}$ relative to molecular hydrogen \citep{Herbst2009}, but we assume that the molecular abundance can be depleted by freeze out at low temperatures or destroyed through reactions.  Assuming that freeze out and thermal desorption are in equilibrium, the threshold density at a given temperature $T$ above which CH$_3$CN will be frozen out is
\begin{equation}
    \rho_{\textrm{thr}} = \frac{m_{\rm d}}{\delta\pi\sigma_{\rm d}} \sqrt{\frac{N_{\rm s}E_{\rm b}}{k_{\rm B}T}}\exp\left(-\frac{E_{\rm b}}{k_{\rm B}T}\right)
\end{equation}
where $\delta$, $m_{\rm d}$, $\sigma_{\rm d}$, $N_{\rm s}$ and $E_{\rm b}$ are the dust-to-gas mass ratio, mean dust grain mass, dust cross section, the number density of surface binding sites ($\approx1.5\times10^{15}$\,cm$^{-2}$) and the species-dependent binding energy $E_{\rm b}/k_{\rm B}=4680$\,K \citep{2004MNRAS.354.1133C}.  We assume a mean grain size of 0.1\,$\mu$m, and grain density of 3.5\,g\,cm$^{-3}$ (from which the cross section and grain mass can be computed assuming spherical grains). We also assume a dust-to-gas mass ratio of $10^{-2}$. Where CH$_3$CN is frozen out we assume a negligible abundance of $10^{-20}$.  We do not account for destruction of CH$_3$CN by two-body reactions/cosmic rays (despite the timescale for destruction by these processes being short) because doing so requires a sophisticated chemical network including CH$_3$CN production pathways. We do however destroy CH$_3$CN in regions where the temperature is above 2300\,K, which is the regime in which the chemical models of \citet{2014A&A...563A..33W, 2015A&A...582A..88W} find its abundance sharply drops and becomes negligible. 

\smallskip

With the abundances and level populations known, a position-position-velocity datacube is constructed using a ray tracing scheme through the disc \citep{2010MNRAS.407..986R}. Both the continuum and line emission are calculated in the frequency range $\nu \pm \nu u_{\rm max}/c $, where $\nu$ is the rest frequency of the molecular line considered in a given synthetic observation, $u_{\rm max}=40$\,km\,s$^{-1}$ (which probes right down to our inner radius of 50\,AU for the 20\,M$_\odot$ star we consider) and $c$ is the speed of light.

\subsection{Accounting for the instrumentational response of ALMA using CASA }
\label{sec:casa}
Given that we are interested in interpreting observations from ALMA, it is essential that we account for the instrumentational response. For example, limited time on source, finite resolution and the lack of sensitivity to large scale structure inherent to high resolution interferometry can all have significant effects on the resulting data.  We therefore postprocess the \textsc{torus} molecular line datacubes described in section \ref{sec:RT}  using the \textsc{casa}\footnote{\url{https://casa.nrao.edu/}} software \citep{2007ASPC..376..127M} to account for the above. 

\smallskip

In our fiducial model we consider band 6 observations in ALMA configuration 40.7, which has a minimum baseline of 81\,m, a maximum baseline of 3.7\,km, an angular resolution of 0.09\arcsec and a maximum recoverable scale of 0.8\arcsec. %\footnote{\url{https://www.iram.fr/IRAMFR/ARC/documents/cycle4/ALMA_Cycle4_Technical_Handbook-Final.pdf}}. 
Given that a 1000\,AU disc at $\sim$\,$1$\,kpc has an angular size of $\sim$\,$1\arcsec$, and given ALMA capabilities in cycle 4, the cycle at which ALMA reached close to full operations for the 12-m array, this configuration was a pragmatic choice for attempting to better detect and resolve discs around massive YSOs and perhaps also to search for substructure within them. We use a spectral resolution of 0.4\,km\,s$^{-1}$, representative of the practical spectral resolution achieved in cycle 4 observations of candidate high mass star-disc systems (Ilee et al., in prep.); however higher spectral resolution is possible with ALMA. 

%his is representative of the practical spectral resolution achieved in cycle 4 observations of candidate high mass star-disc systems.  

\smallskip

To account for the instrument response of ALMA we use the \textsc{casa} \textsc{simobserve} and \textsc{clean} routines. We set the zenith precipitable water vapor to $1.796$\,mm (an estimate from the ALMA sensitivity calculator for the fifth octile). The total time on source is set to 2 hours (with an integration time of $20$\,s), with the source transiting in the middle of the observation. We obtain an image of the continuum emission from the 40 line-free channels at either end of the cube. Thus, the continuum bandwidth used here is $\sim$\,0.025\,GHz.  Even in line-rich sources such as NGC6334I, line-free continuum bandwidths $\gtrsim$10$\times$ larger are readily obtainable with ALMA \citep[see e.g.\ Table 1 of][]{hunter_2017}, so our results represent a conservative lower limit for continuum signal-to-noise ratios. The line-free channels are used to subtract the continuum in the \textit{uv}-plane, using the \textsc{uvcontsub} routine. To produce the continuum and line images we use the \textsc{casa} \textsc{clean} routine with a threshold of 3$\times$ the RMS noise (RMS $\approx 2.5$\,mJy beam$^{-1}$ per channel for the K=3 line, and $\approx 0.3$\,mJy beam$^{-1}$ for the continuum), and using a circular mask around the source. Integrated intensity (moment 0) and intensity-weighted velocity (moment 1) maps are made using the \textsc{immoments} routine, and the position-velocity (PV) diagrams are made using the \textsc{impv} routine. In making moment 1 maps, only pixels above a cutoff of 5$\times$ the RMS are included.

%This is a rather narrow band-width of 
%40*0.4 km/s 
%	= 16 km/s 
%	~ 16/3e5 x 239 GHz 
%	~ 0.01GHz 

%In TDM (time domain mode) with ALMA, you can easily get a few GHz of continuum bandwidth in addition to lines.  %Hence, the SNR here are likely significantly underestimated with regards to a real observation.  

%I would quantify the bandwidth that you have here with the caveat that significantly larger bandwidths will result in much larger SNR in a real observation for the continuum.

\smallskip

We adopt a distance of 3\,kpc for the majority of our models in this paper, but also explore 1-5\,kpc. We are hence using substantially larger distances than previous synthetic observations of discs about massive YSOs, which typically consider 0.5-1\,kpc and/or note that at larger distances detecting discs and structure is difficult \citep[e.g. the 2\,kpc distant models of][]{2007ApJ...656..959K}.  In all observations considered here, we assume our object lies in the direction of the centre of the protocluster G11.92$-$0.61, with on sky co-ordinates of R.A. = 18h13m58.1, Dec. = $-$18d54m16.7s \citep{cyganowski_2017}.

\section{Results}
\label{sec:results}

We use our framework for quickly producing synthetic observations of self-gravitating discs to explore whether or not self-gravitational substructure, spiral density waves and fragmentation, are detectable in discs around massive stars with upcoming ALMA observations. We find that the spiral features are not always obvious neither spatially, in continuum images and integrated intensity (moment 0) maps of line emission, nor spectrally, in intensity-weighted velocity (moment 1) maps and position-velocity (PV) diagrams. This is in contrast, but not in conflict, with the findings of \cite{2013MNRAS.433.2064D},  who modelled signatures of self-gravitating discs in \textit{nearby, lower mass} systems (their model is of a 1\,M$_\odot$ star and a 0.39\,M$_\odot$ disc) in lines from more abundant species such as CO and HCO$^+$. Significant enhancement of substructure can, however, be obtained, using filtering techniques presented in this section.

\smallskip

Our fiducial model is a disc around a $M_*=20$\,M$_\odot$ star at a distance of $3$\,kpc with an inclination of $30^\circ$, an accretion rate of $\dot{M}=10^{-3}$\,M$_\odot$\,yr$^{-1}$, a ratio of atmospheric to mid-plane temperature $f_{\rm atm}=4$, an intensity of velocity and temperature perturbations $\eta=0.2$, and two spiral arms (see the second row of Fig. \ref{discplot}). In section \ref{sec:discussion} we will explore how the detection of self-gravitational signatures is sensitive to the disc and observational (e.g. distance, line, inclination) parameters. In this section we focus on enhancing substructure using filtering techniques.

\subsection{Enhancing spatial detection of substructure in discs}
\label{sec:spatial_filtering}

As mentioned above, spirals are not always obvious in our synthetic continuum images and moment 0 maps. To improve the spatial detectability of spirals we subtract the continuous background disc emission to highlight the more spatially confined substructure. We define the base synthetic image (i.e. that from \textsc{torus} and \textsc{casa}) as the ``primary image''. We consider subtraction of 2D (elliptical) Gaussian fits to this primary image using the \textsc{casa} \textsc{imfit} routine. The \textsc{imfit} routine is a well-tested, easy-to-use component of the publicly available \textsc{casa} software.

\smallskip

\begin{figure*}
    \centering
    \includegraphics[scale=0.2,trim={0 0 5.5cm 0},clip]{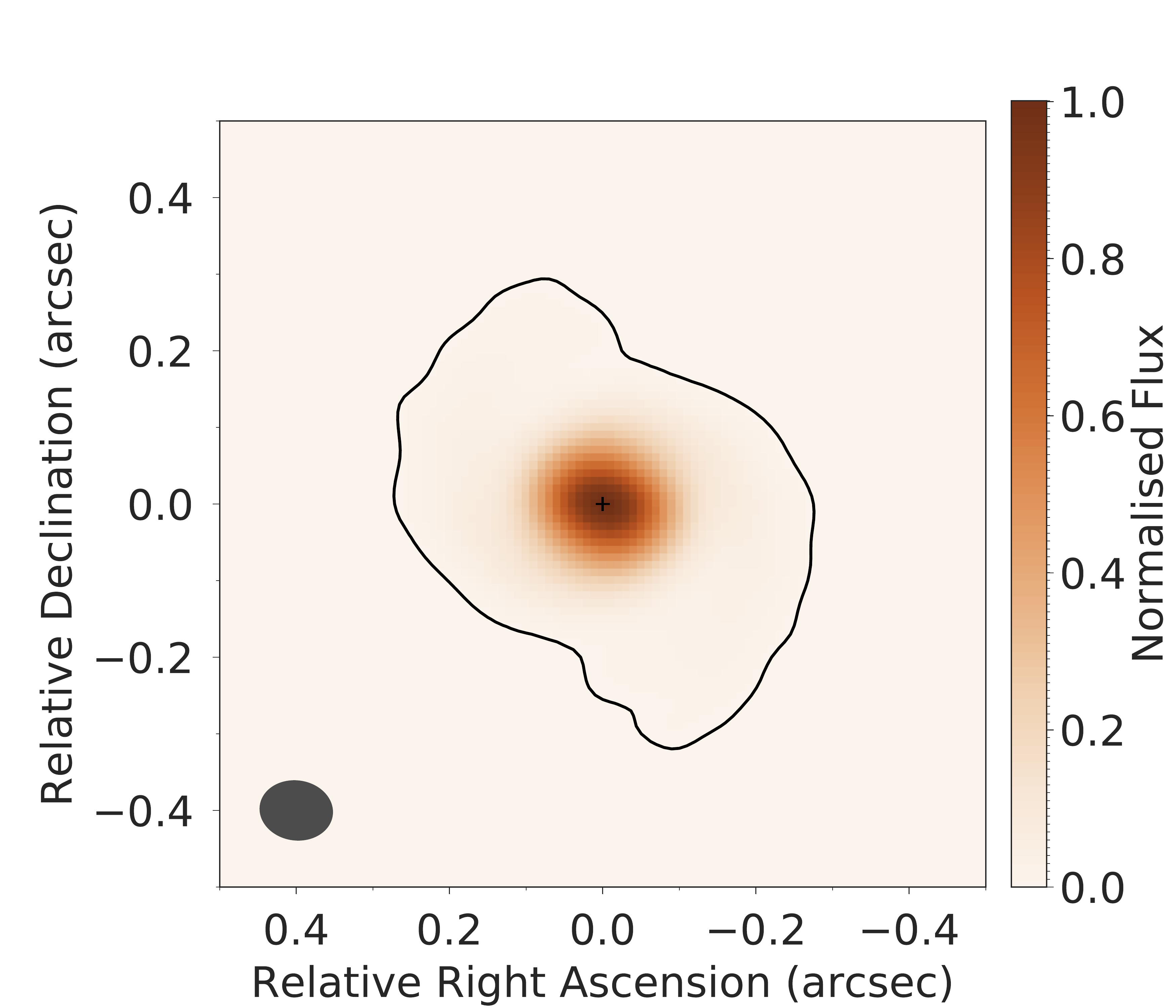}
    \includegraphics[scale=0.2,trim={6cm 0 5.5cm 0},clip]{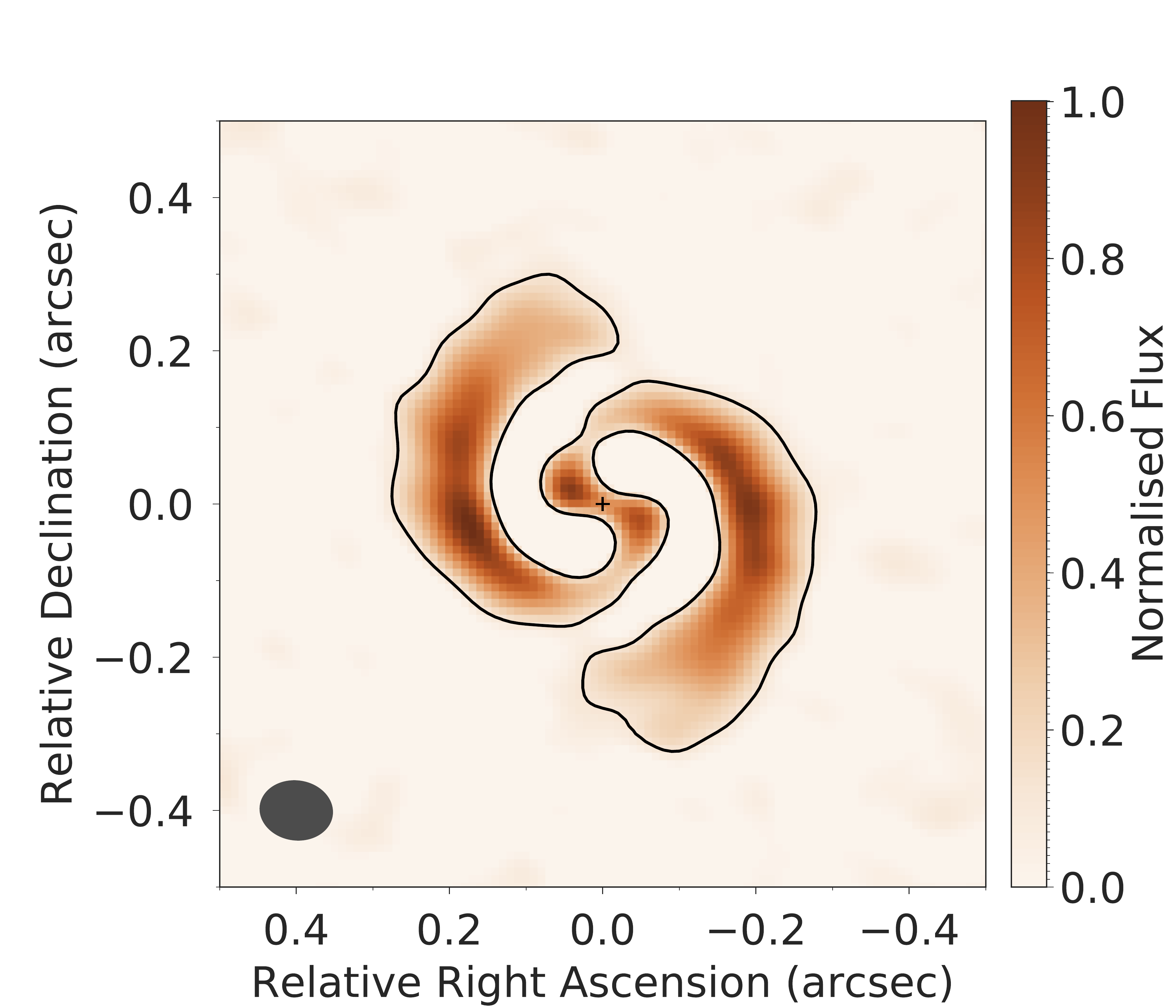}
    \includegraphics[scale=0.2,trim={6cm 0 0 0},clip]{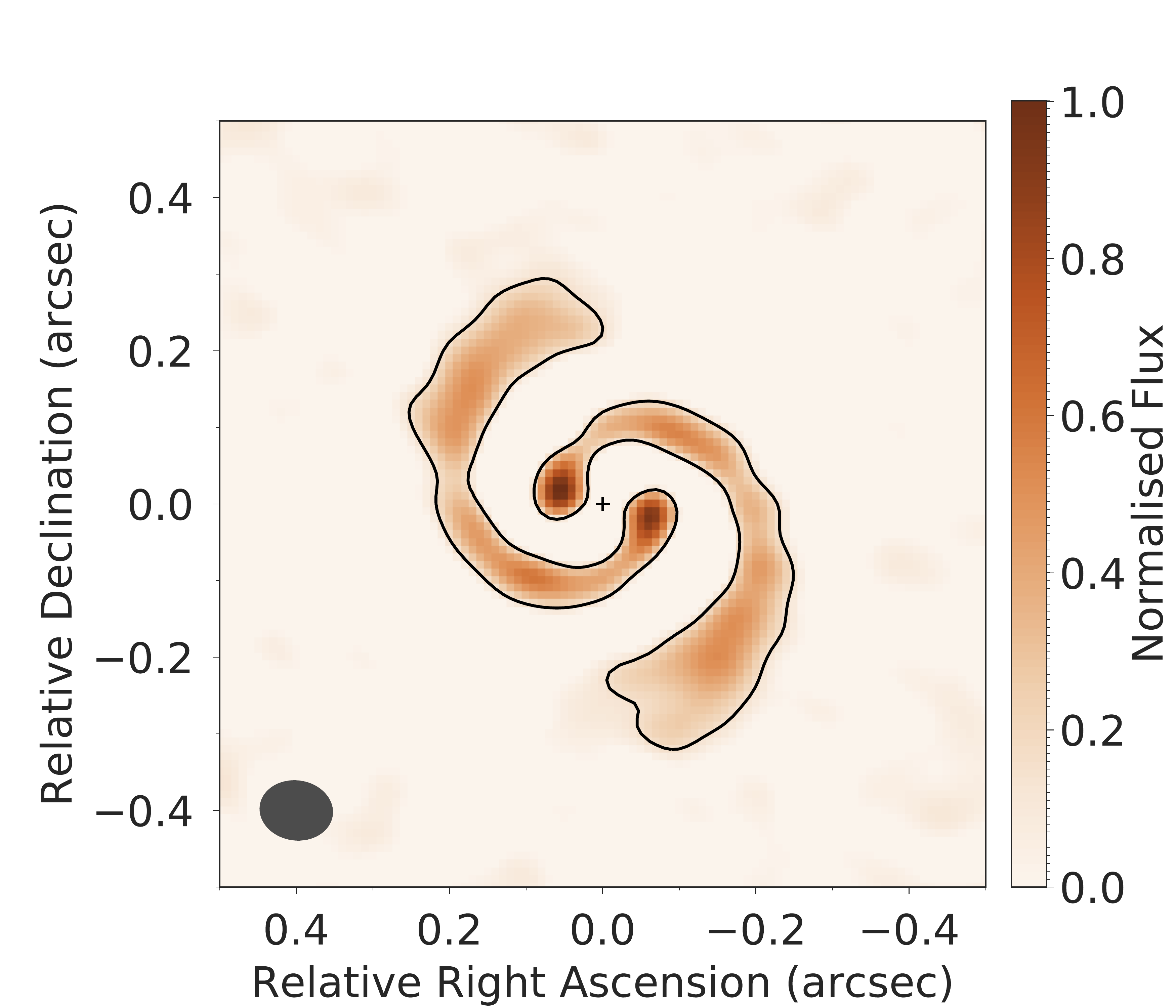}
    \caption{Synthetic continuum images for a 2-arm spiral disc model with two filtering techniques applied, with $3\sigma$ ($\sigma=$\,off-source RMS noise) contours. Flux is normalised with respect to the peak value in each image. Left is the primary (unfiltered) image of the 239\,GHz continuum, with a signal-to-noise ratio of 500. Middle subtracts a single Gaussian fit to the primary image. Right subtracts a two-component Gaussian fit. The signal-to-noise ratio in the filtered images is 25 and 20, for the single and double Gaussian fits respectively.}
    \label{fig:subtractionContinuum}
\end{figure*}

\begin{figure*}
    \centering
    \includegraphics[scale=0.2,trim={0 0 5.5cm 0},clip]{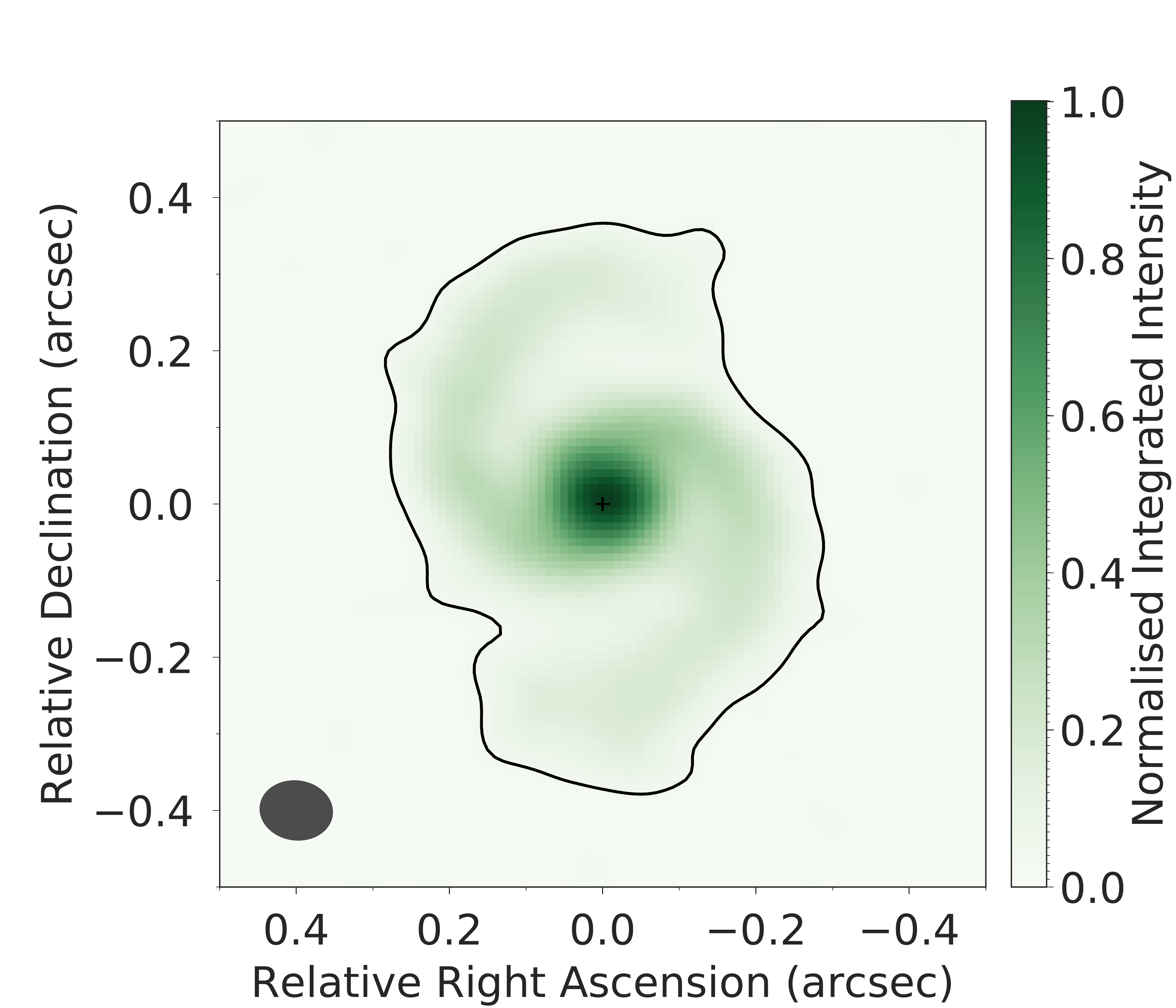}
    \includegraphics[scale=0.2,trim={6cm 0 5.5cm 0},clip]{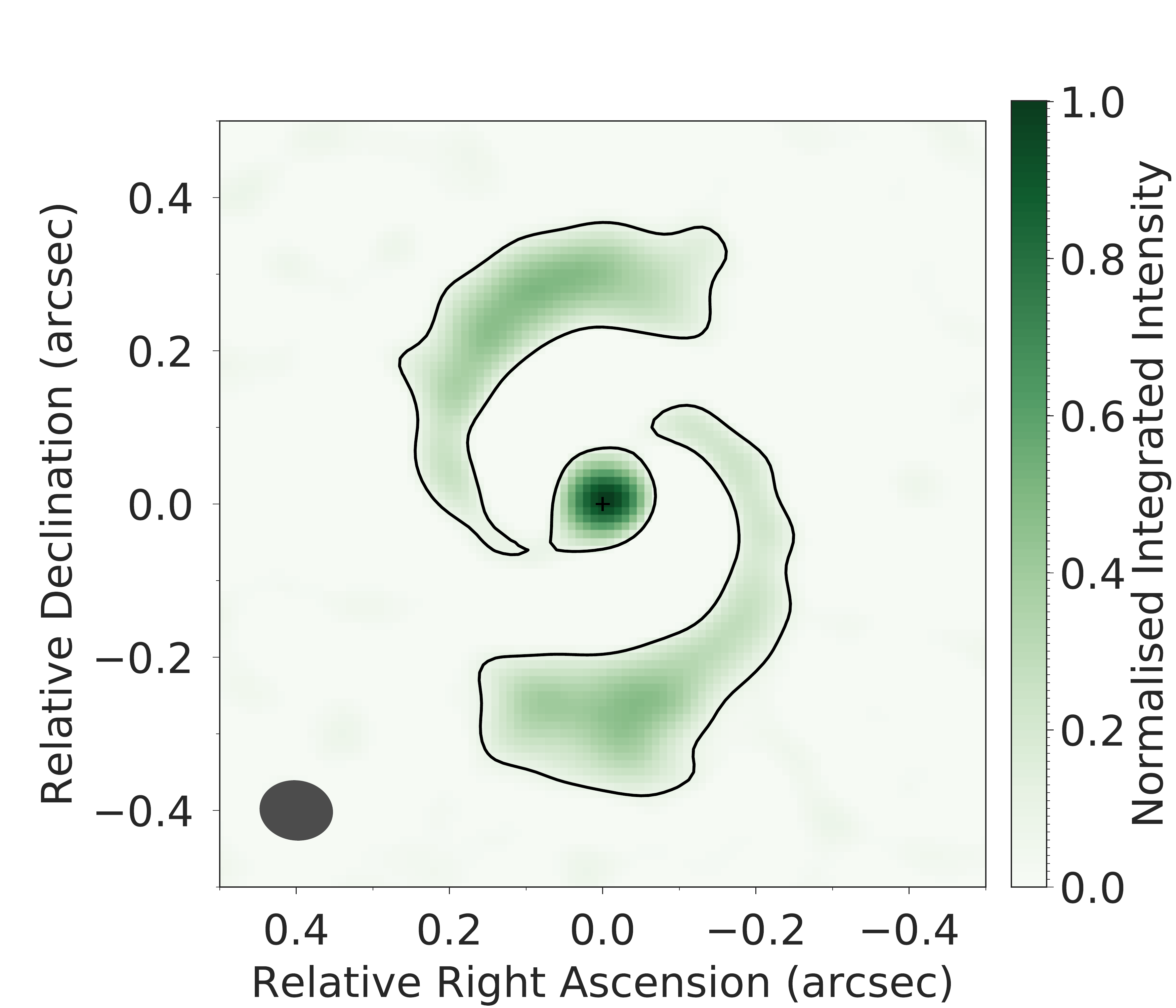}
    \includegraphics[scale=0.2,trim={6cm 0 0 0},clip]{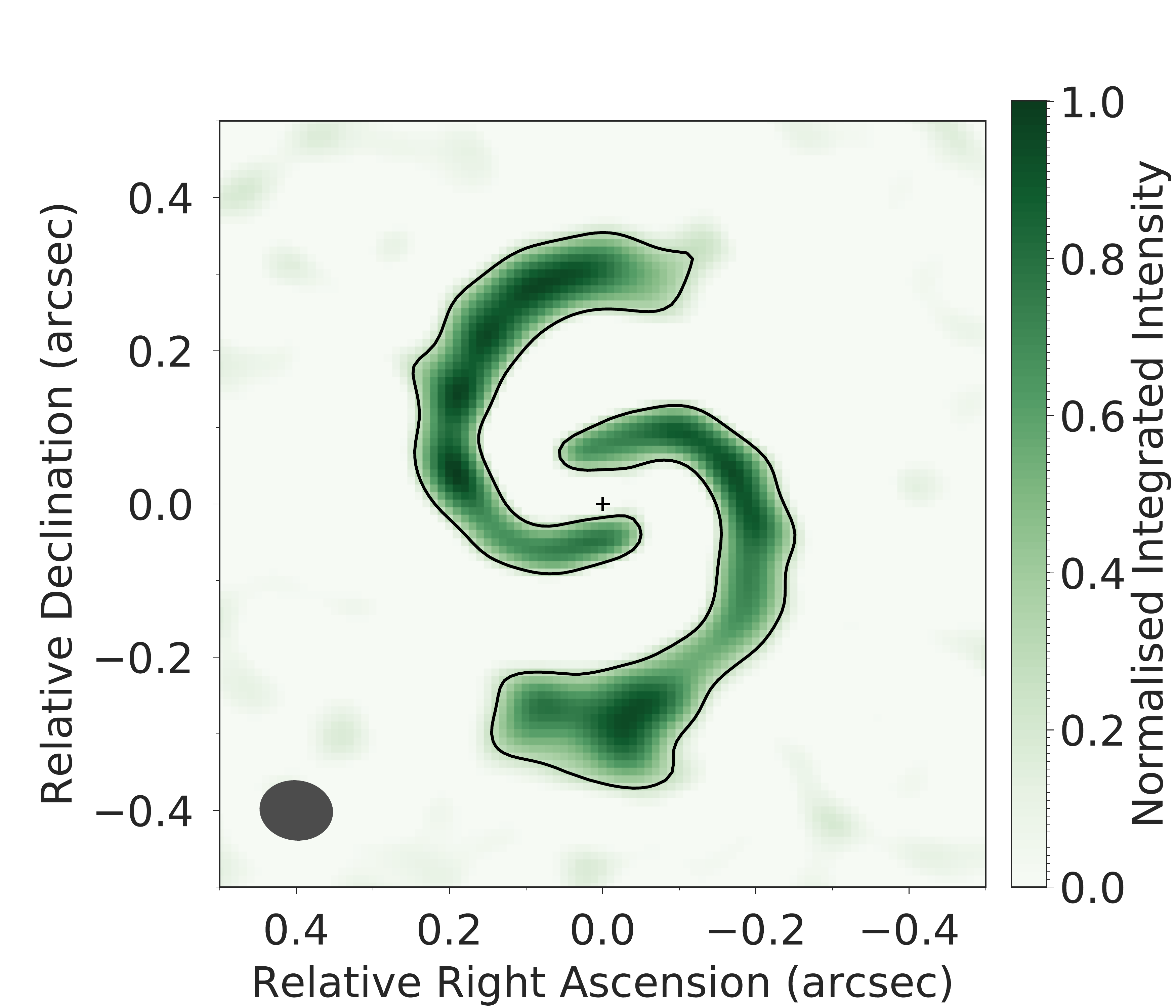}
    \caption{Synthetic moment 0 maps for a 2-arm spiral disc model with two filtering techniques applied, with $3\sigma$ ($\sigma=$\,off-source RMS noise) contours. Integrated intensity is normalised with respect to the peak value in each map. Left is the primary (unfiltered) image, with a signal-to-noise ratio of 82. Middle subtracts a single Gaussian fit to the primary image. Right subtracts a two-component Gaussian fit. Signal-to-noise ratio in the filtered images is 33 and 11, for the single and double Gaussian fits respectively.}
    \label{fig:subtractionMom0}
\end{figure*}

Figure \ref{fig:subtractionContinuum} compares the primary image with the residuals of fitting a single Gaussian and of simultaneously fitting  two Gaussian components to the 239\,GHz continuum image of our fiducial disc model with two spiral arms.  Figure \ref{fig:subtractionMom0} presents the same comparison for  moment 0 maps of the line emission. We find that in the moment 0 maps the two-component Gaussian fit is better at filtering the more continuous disc emission, as one component fits the bright inner disc (that is optically thick, see section \ref{sec:optical_depth}) and the other covers a larger extent of the disc. On the other hand, subtraction of a single Gaussian fit from the moment 0 map of our fiducial disc model (the middle panel of Fig. \ref{fig:subtractionMom0}) is inadequate, as it does not fit the continuous/axisymmetric emission from the disc well. It does not subtract the unresolved bright inner disc, but subtracts the inner regions of the spiral arms that are clearly revealed in the image filtered using the two-component Gaussian fit. On the other hand, we find that in the continuum image it is sufficient to fit and subtract a single elliptical Gaussian profile to clearly highlight the spiral substructure. Note that the continuum image filtered using the two-component Gaussian fit (right hand panel of Fig. \ref{fig:subtractionContinuum}) features two compact bright spots near the center of the disc that are not real structures. The artefacts also appear in the image filtered using a single Gaussian profile, but they are much less significant.

\smallskip

Throughout the rest of this paper we apply the technique of subtracting the one-component and the two-component Gaussian fits when considering continuum and moment 0 line emission maps respectively. We found this technique to have a good balance of ease of implementation and ability to detect substructure.

\smallskip

We note two points of caution when using these techniques. Firstly, our results show that even in discs that do not possess real clump-like features (such as our models with completely smooth spirals), the interferometric imaging can induce artefacts that could be mistaken for clumps and/or substructure in the disc. We therefore suggest that caution be taken when attributing such features to the presence of a real feature, such as disc fragmentation due to gravitational instability, particularly when the feature is of low significance and close in spatial scale to the interferometric beam. Secondly, the \textsc{casa} \textsc{imfit} routine requires an initial estimate of the parameters for multiple component Gaussian fits as input, and the results can be sensitive to this input. We find it useful to run the fitting routine several times with varying input estimates to check for convergence. 

\smallskip

Finally, we note that fitting more tailored parametric axisymmetric disc models of the intensity profile (e.g., those presented here) could do an even better job at revealing substructure, particularly since the brightness profile will deviate from Gaussian in practice. However this requires more detailed modelling.

\subsection{Enhancing spectral detection of substructure in discs}
\label{sec:kinematic_filtering}

To improve the kinematic detectability of spirals we subtract line-of-sight projected Keplerian velocities, convolved with the synthesized beam of our synthetic observations from our model moment 1 maps. This method was used by \citet{2017arXiv171000703W} to constrain the inclination and the position angle of the HD 100546 disc, as well as to infer the presence of either a warp or radial flow  residing $< 100$\,AU from the central star \citep[distinguishing between the two is difficult, e.g.][]{2014ApJ...782...62R}.

\smallskip

By using the same fitting metrics as \citet{2017arXiv171000703W} we find that the best fit inclination and position angle do not always correspond to the known (i.e. input) inclination and position angle of our disc models, with typical variation in both of $\pm 10$\,degrees. We attribute this to deviations from a simple flat thin disc model. Irrespective of this, we find that an incorrect best fit inclination (that is within $\pm 5$--$15$\,degrees from the correct value) does not affect the ability to detect the spiral signatures in our synthetic observations nor does it induce the appearance of any artificial substructure. We thus proceed with subtraction of Keplerian profiles corresponding to the actual disc model inclinations and position angles, but noting that the automatically fitted value would also suffice. 

\smallskip

\begin{figure*}
    \centering
    \includegraphics[scale=0.18,trim={0 0 5.5cm 0},clip]{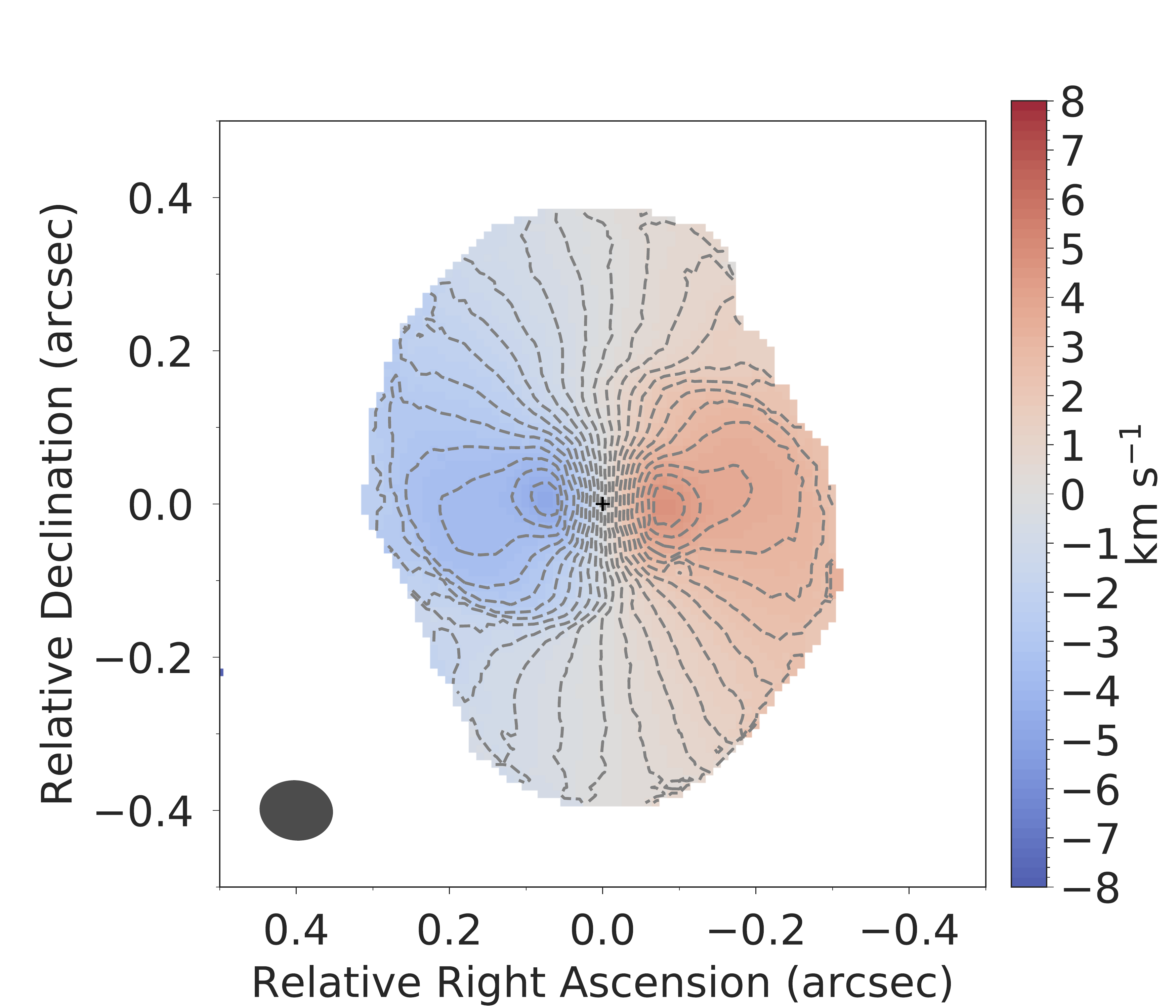}
    \includegraphics[scale=0.18,trim={6cm 0 0 0},clip]{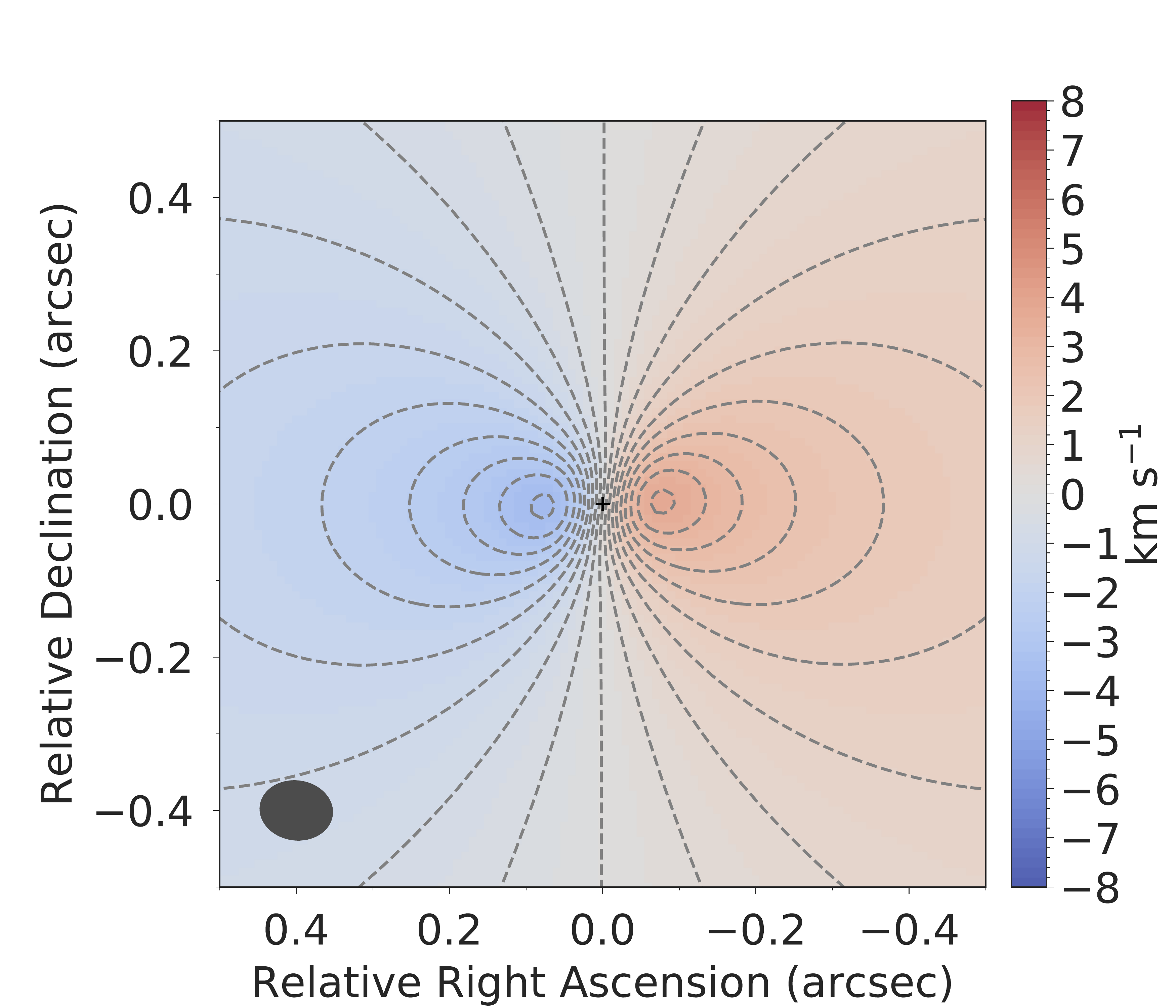}
    \includegraphics[scale=0.18,trim={6cm 0 0 0},clip]{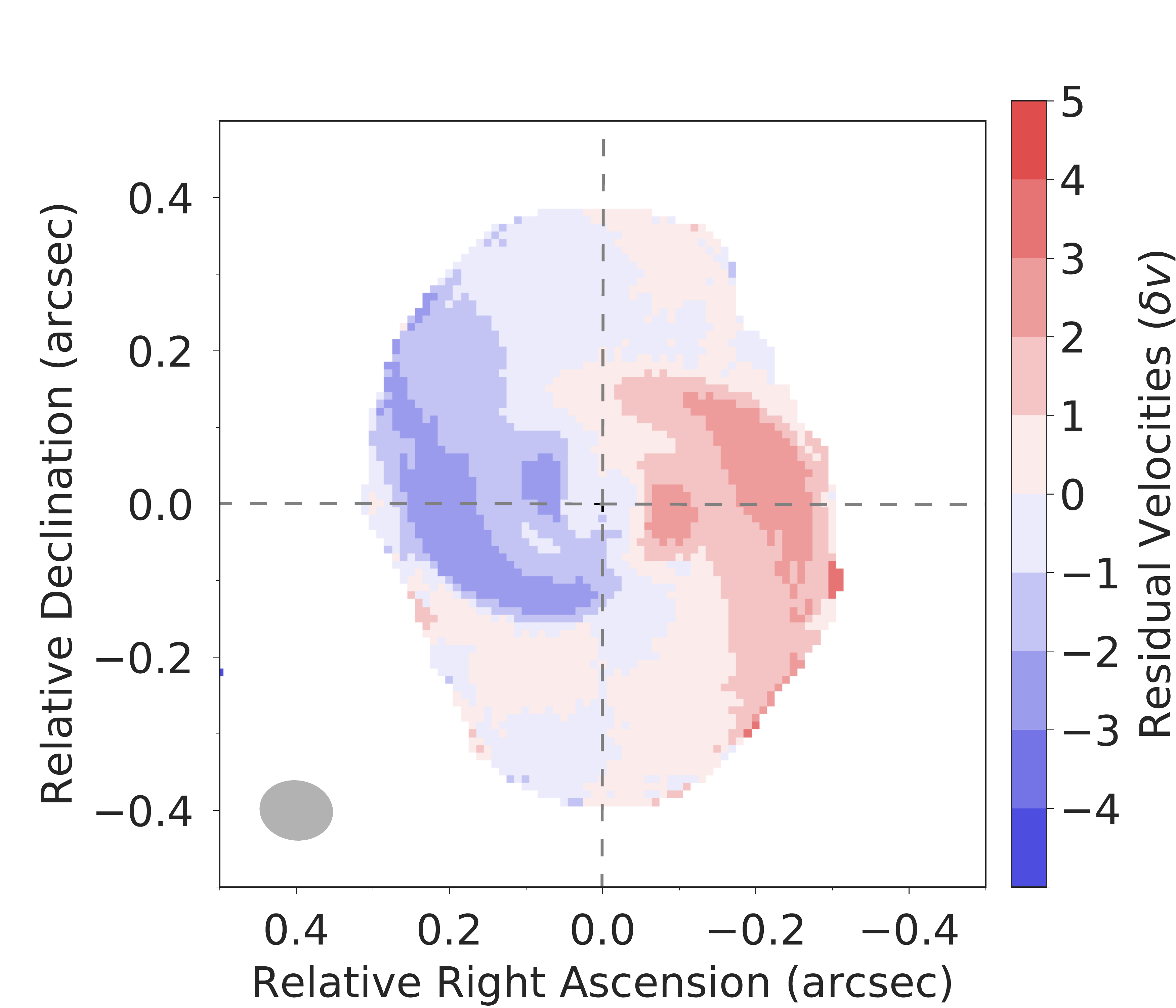}
    \caption{An illustration of kinematic filtering applied to a moment 1 map of a 2-spiral disc. The left hand panel is the initial synthetic observation, the middle a Keplerian profile and the right hand the residual. The velocity residuals are given in units of the spectral resolution $\delta v=0.4$\,km\,s$^{-1}$. Dashed contours in the left hand and the middle panel also show the velocity in units of $\delta v$. In the right hand panel dashed horizontal and vertical lines indicate disc position angle and the line perpendicular to it, respectively. The warp-like features resulting from spirals are much more evident in the residuals map.}
    \label{fig:subtractionMom1}
\end{figure*}

The moment 1 map of the synthetic line emission observation, the moment 1 map of equally inclined flat Keplerian emission, and the result of the subtraction of the latter from the former map (the moment 1 map residuals) are shown in Fig. \ref{fig:subtractionMom1} for our fiducial disc model with two spiral arms. The velocity residuals are given in units of the spectral resolution $\delta v=0.4$\,km\,s$^{-1}$. The warp-like features resulting from spiral structure are much more prominent following the kinematic filtering technique we apply. 

\smallskip

Throughout the rest of this paper we apply the described subtraction technique to enhance the kinematic detectability of disc substructure in moment 1 maps and present only the residuals of the subtraction.

\section{Discussion}
\label{sec:discussion}

The speed of our disc model construction and synthetic observations (under an hour per model using four cores for the radiative transfer calculations, and a single core for the rest, on a desktop) allows us to explore a parameter space of disc (e.g. mass, temperature) and observing (e.g. distance, line) parameters. We detail the specific parameter space that we explore in Table \ref{tab:parameter_space}. Starting from our fiducial disc model, we vary the parameters one-by-one, producing sets of models that are different from each other in the value of a single parameter. This is with the exception of disc inclination which is also separately set for each set of models. All of the models are available for download\footnote{DOI 10.5281/zenodo.1408072.}. In this section we discuss what the limiting factors are for the detection of self-gravitational substructure both spatially and spectrally, and gauge the expectations for upcoming ALMA observations of massive young stellar objects. We focus first on self-gravitating disc models with spiral density waves, for which we also discuss the impact of disc thermal and chemical structure (Section \ref{sec:thermochemical_structure}) and optical depth (Section \ref{sec:optical_depth}). We then explore detectability of disc fragmentation into clumps as a function of the fragment properties (Section \ref{sec:fragment}).

\begin{table*}
    %\thispagestyle{empty}
    %\noindent
    \caption{A summary of the models with spiral arms presented in this paper, except for the models with clumps. The first column lists the model parameters which were varied, second the fiducial (default) value of each parameter, third the values that were explored for each parameter, fourth the disc inclination used for the set of models in each row. In the first column $\eta$ is the magnitude of spiral velocity and temperature perturbation, and $f_{\rm atm}$ the ratio of atmospheric to mid-plane temperature.}
    \label{tab:parameter_space}
    \label{table:modelParams}
    \begin{tabular}{ l c c c c c c c c c } %c}
        \hline
        Parameter & Fiducial & Values explored  & Inclination $(^\circ)$ \\ %& Model IDs \\
        
        \hline
        \hline
        
        Line choice	&	K = 3, J=13 $\rightarrow$ 12	&	K = 7, J=13 $\rightarrow$ 12		&	30	\\ %&	LIN 2\\
        Inclination $(^\circ)$	&	30, 60	&	0\,--\,90		&	\\ %&	INC [N]	\\
        Number of spirals	&	2	&	0, 4		&	30	\\ %&	NSP 0,4 \\
        Accretion rate $(M_{\odot}\rm{yr}^{-1})$ &	$10^{-3}$	&	$10^{-5}$, $10^{-4}$, $10^{-2}$		&	60	\\ %&	ACC 1,2,3 \\
        $\eta$	&	0.2	&	-0.2, 0.1		&	60 \\ %&	ETA 1,2	\\
        Rotational offset (rad)	&	0	&	$\pi/4$, $\pi/2$, $3\pi/4$	&	60	\\ %&	ANG [N] \\
        Distance (kpc)	&	3	&	1, 5	&	30 \\ %&	DIS 1,2		\\
        $f_{\rm atm}$	&	4	&	1, 2, 8		&	60	\\ %&	ATM 1,2,3\\
        Dust	&	$a_{\rm max}=1$\,cm, $q=3.3$	&	$a_{\rm max}=0.1$\,$\mu$m\,$-1$\,cm;  $q=3.3-3.5$	&	60	\\ %&	DUST 1, 2 \\
        \hline
    \end{tabular}
\end{table*}

\subsection{Spatially resolving spiral substructure}

\begin{figure*}
    \centering
    \includegraphics[scale=0.18,trim={0 0 5.5cm 0},clip]{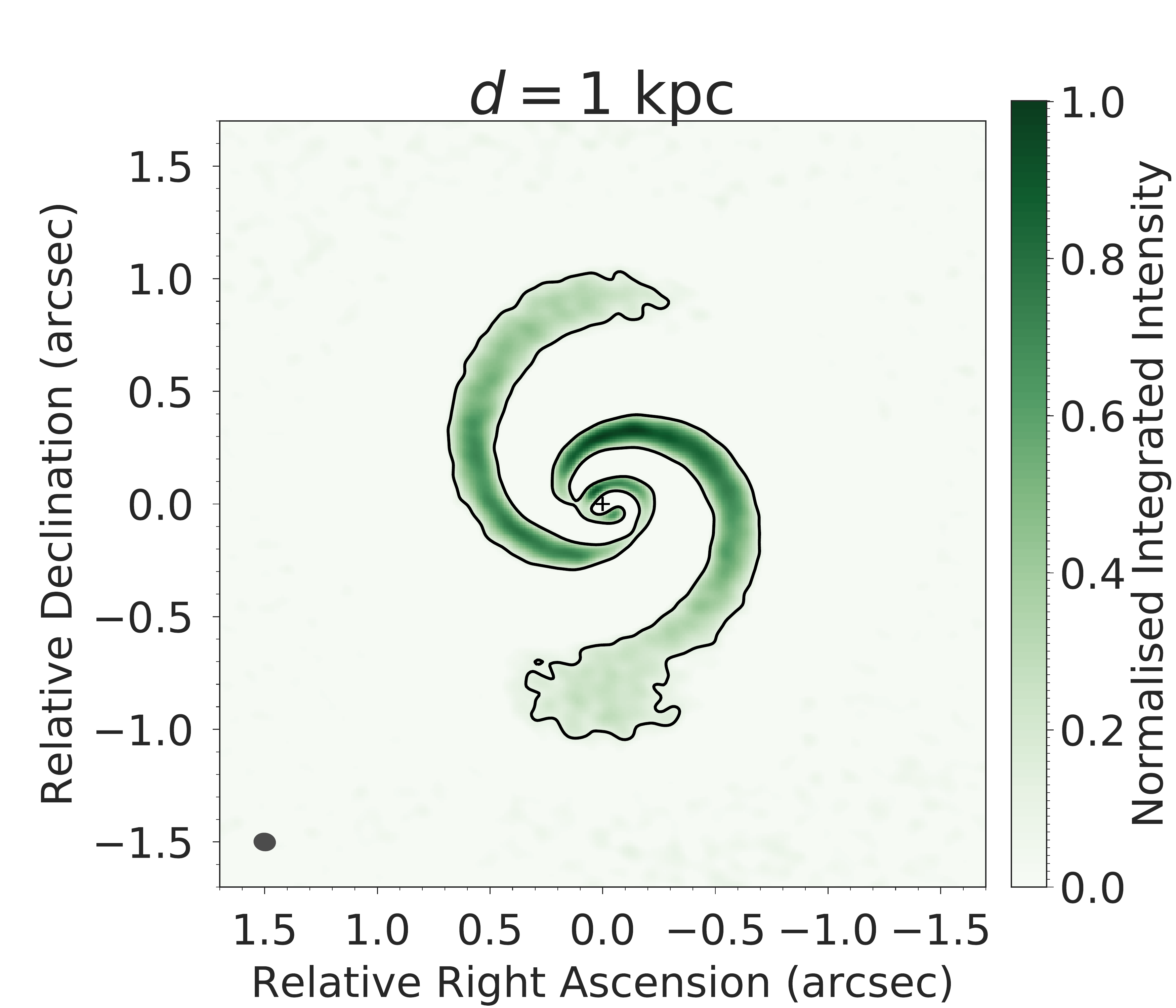}
    \includegraphics[scale=0.18,trim={0 0 5.5cm 0},clip]{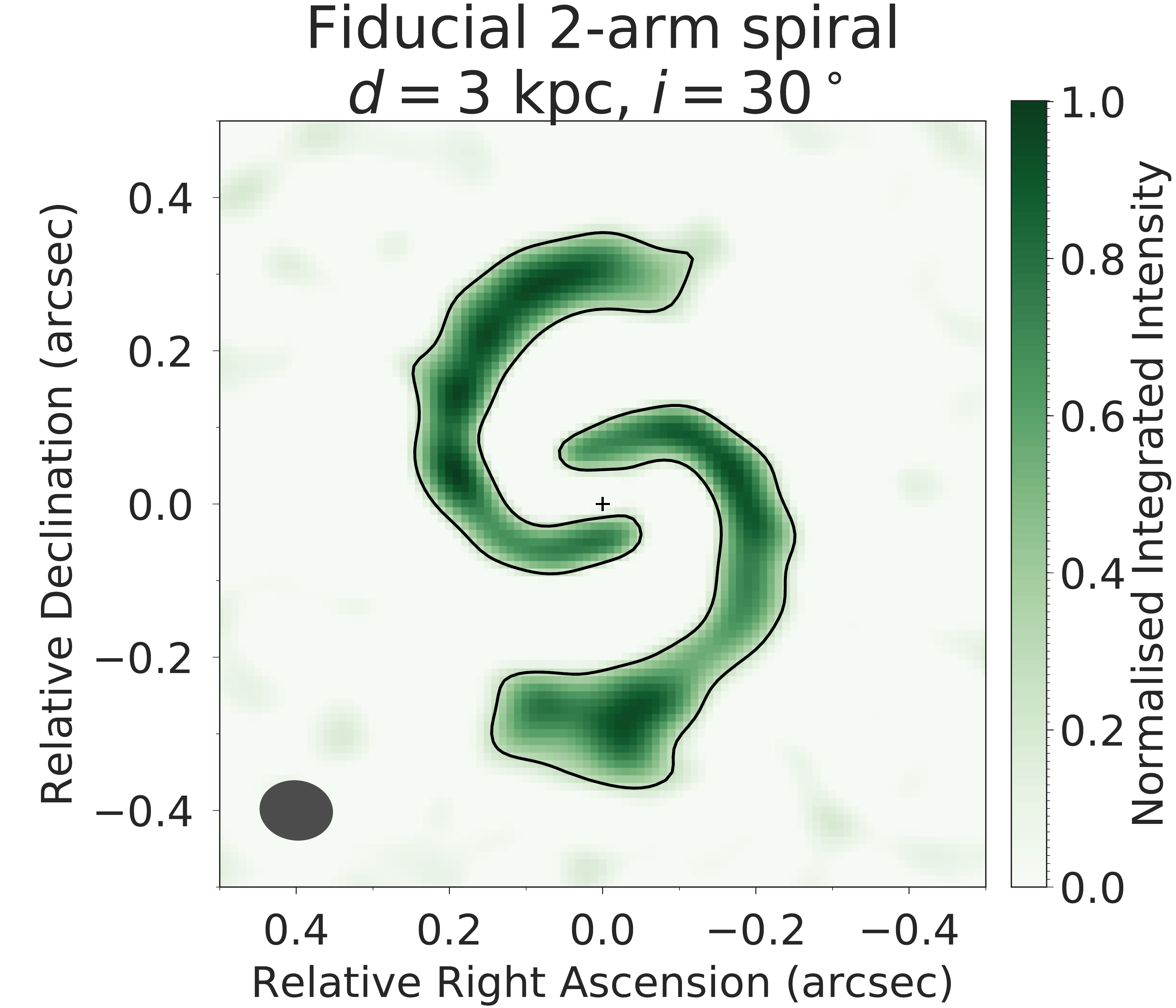}
    \includegraphics[scale=0.18,trim={6cm 0 0 0},clip]{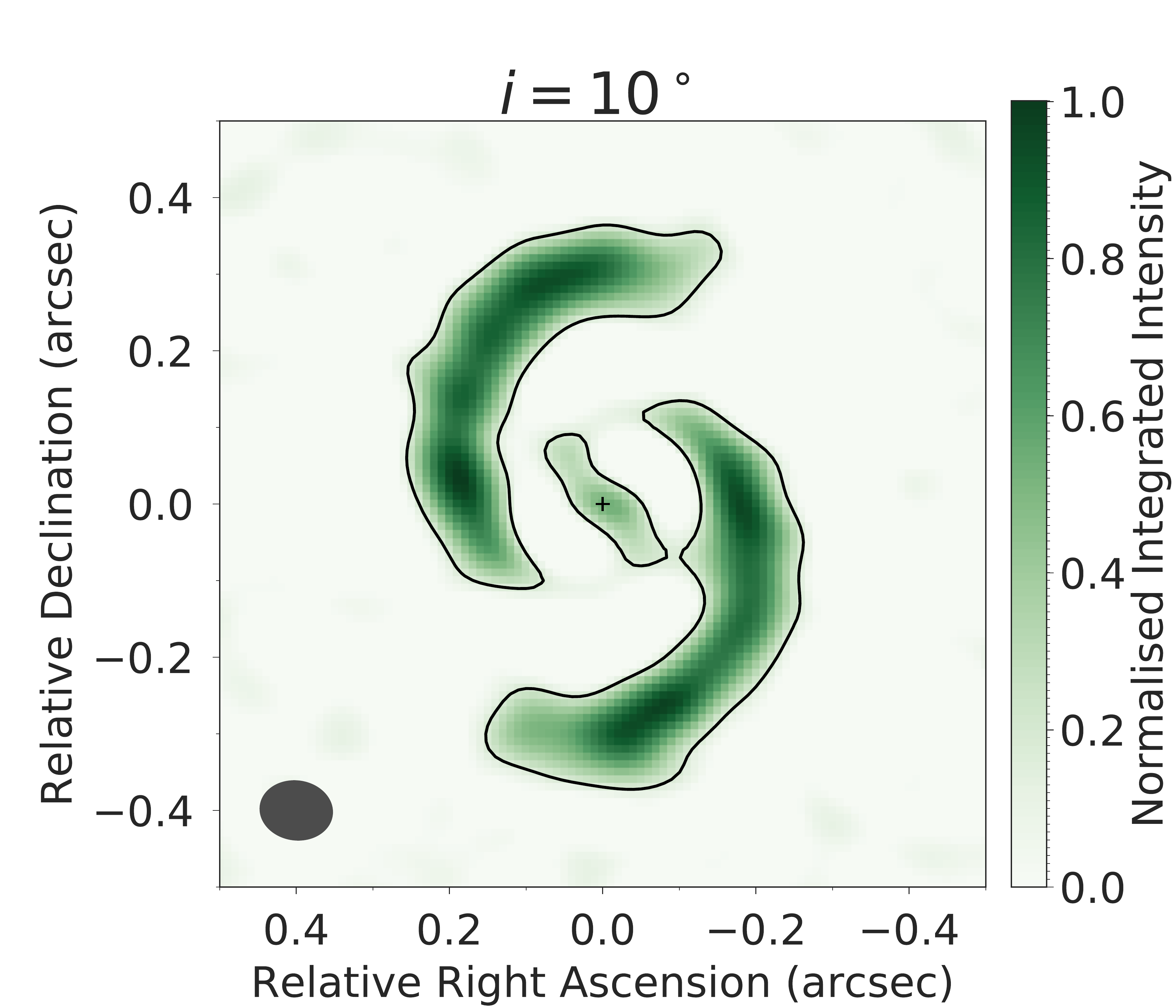}
    \\
    \includegraphics[scale=0.18,trim={0 0 5.5cm 0},clip]{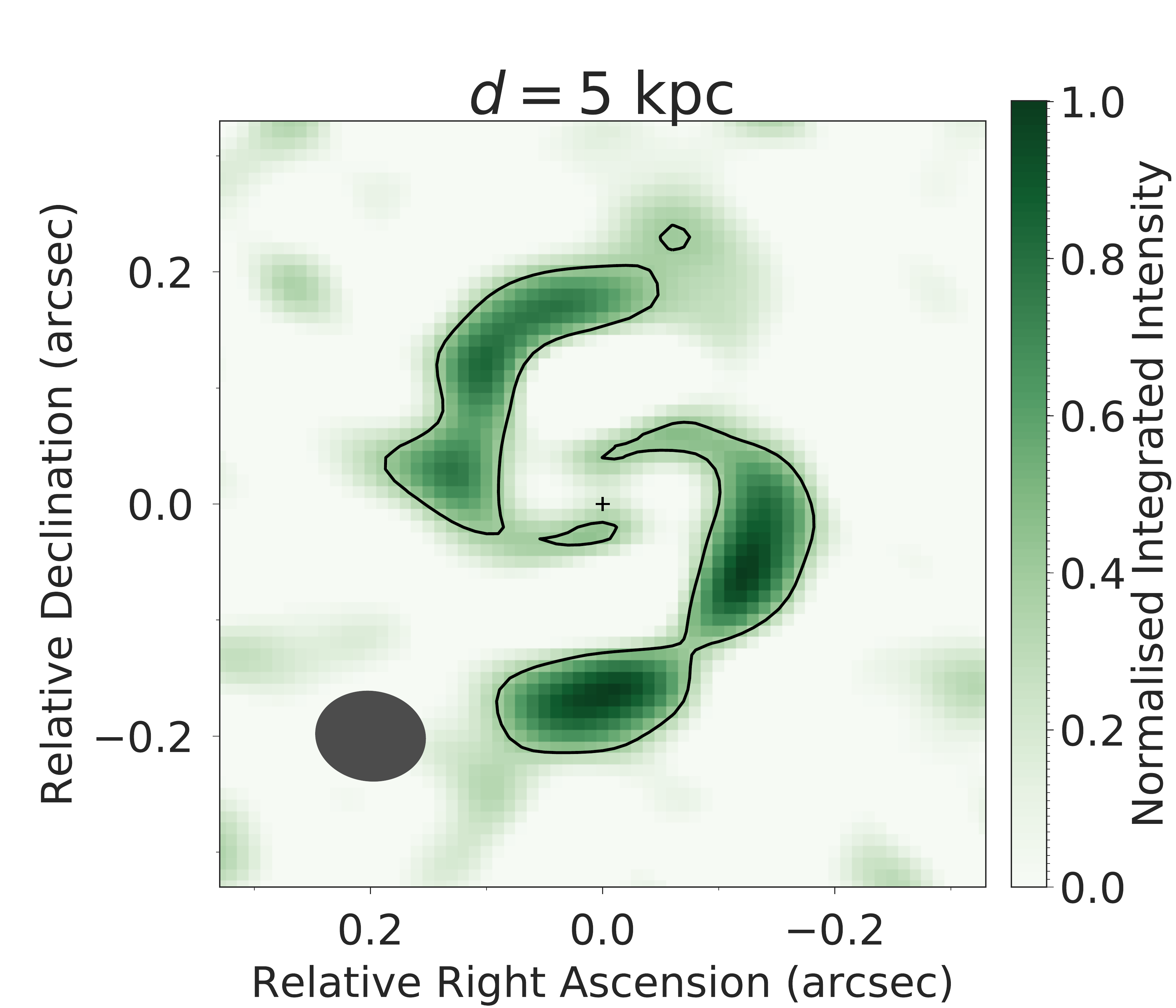}
    \includegraphics[scale=0.18,trim={0 0 5.5cm 0},clip]{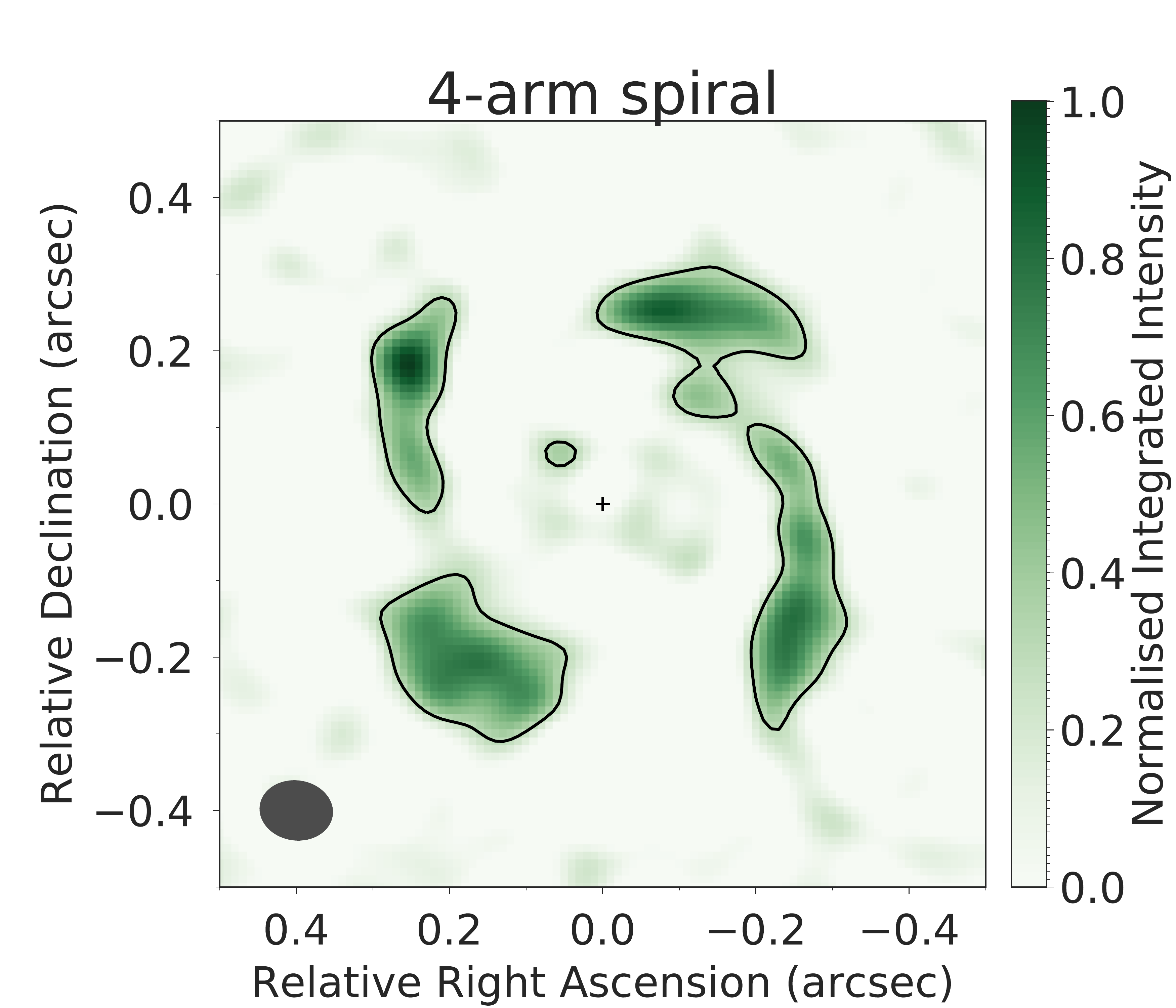}
    \includegraphics[scale=0.18,trim={6cm 0 0 0},clip]{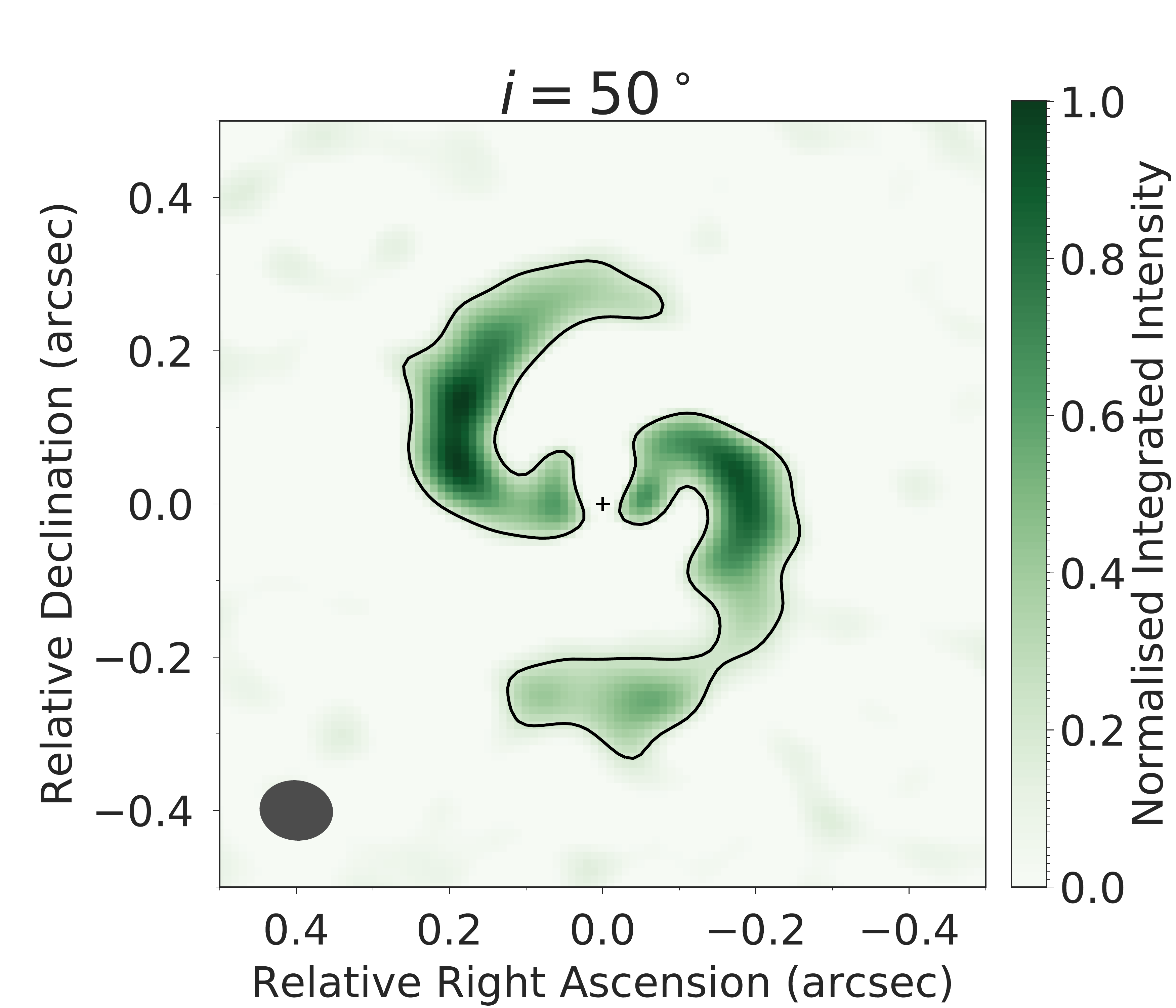}
    \caption{A summary of the parameters to which the spatial detection of spirals is sensitive. All of the panels show line moment 0 maps from synthetic observations, with $3\sigma$ ($\sigma=$\,off-source RMS noise) contours, and in which 2-component Gaussian fits are subtracted to enhance substructure (see section \ref{sec:spatial_filtering}). For the six moment 0 maps shown here, the signal-to-noise ratio in the unfiltered images varies roughly between 55 and 110, and in the filtered images between 8 and 25; the lower boundaries of these ranges correspond to the disc model at a distance of $d=5$\,kpc, and the upper boundaries to the disc model at $d=1$\,kpc. Integrated intensity is normalised with respect to the peak value in each map. The top central panel is our fiducial 2-arm spiral model. The lower central panel has 4 spirals, but is otherwise identical to the 2-arm fiducial. The left hand panels illustrate the sensitivity to distance and the right hand panels the sensitivity to inclination.}
    \label{fig:spatialVariations}
\end{figure*}

Spatial detection of spiral substructure is sensitive to the angular resolution relative to the angular separation between spirals, as viewed by the observer. Manifestly, the spatial detectability of spirals depends on the disc distance, as well as on the disc inclination and the nature of spirals, as illustrated in Fig. \ref{fig:spatialVariations}. This section focuses on the line moment 0 maps from our synthetic observations, but in most cases the same conclusions can be drawn from analysis of the continuum images.

\smallskip

Spiral substructure in strongly self-gravitating discs, i.e. those with a lower number of spiral arms and less tightly wound spirals, is likely to be detected. For $1000$\,AU-radius discs with an inclination of 30 degrees, we find that 2-arm spirals are resolvable up to approximately $5$\,kpc distances with ALMA configuration 40.7.  Such a distance encompasses approximately 70 per cent of the massive YSO population in the Red MSX Source (RMS) Survey\footnote{\url{http://rms.leeds.ac.uk}} for which distances are available \citep{lumsden_2013}.  At close distances ($<1$\,kpc) the moment 0 maps of line emission might even yield hints of vertical stratification in the outer disc (top left panel of Fig. \ref{fig:spatialVariations}), since the beam size is smaller than the disc scale height inside the spiral arms. Indeed, the vertical stratification of molecular emission in more evolved, axisymmetric discs has already been observed (e.g. HD 163296, \citealt{rosenfeld_2013, deGregorio-Monsalvo_2013}), but our results suggest that similar structures will be observable in non-axisymmetric discs. 

\smallskip

As the number of spiral arms increases, however, and spirals become more tightly wrapped, it becomes increasingly more difficult to resolve the spirals. We find that 4-arm spirals are barely discernible at a moderate distance of $3$\,kpc. This principally happens because the size of the synthesized beam is large compared to the angular separation between spiral arms.

\smallskip

Spiral substructure is more easily resolved spatially for more face-on discs, but spirals in our models are detected for the majority of disc inclinations, from 0 (face-on discs) up to $\sim$\,$50^\circ$. Furthermore, over the inclinations that the spirals are not spatially detected, below we will show that they are likely to be detected spectrally.

\subsection{Spectrally resolving spiral substructure}
\label{sec:spectrally_resolving}

\begin{figure*}
    \centering
    \includegraphics[scale=0.18,trim={0 0 5.5cm 0},clip]{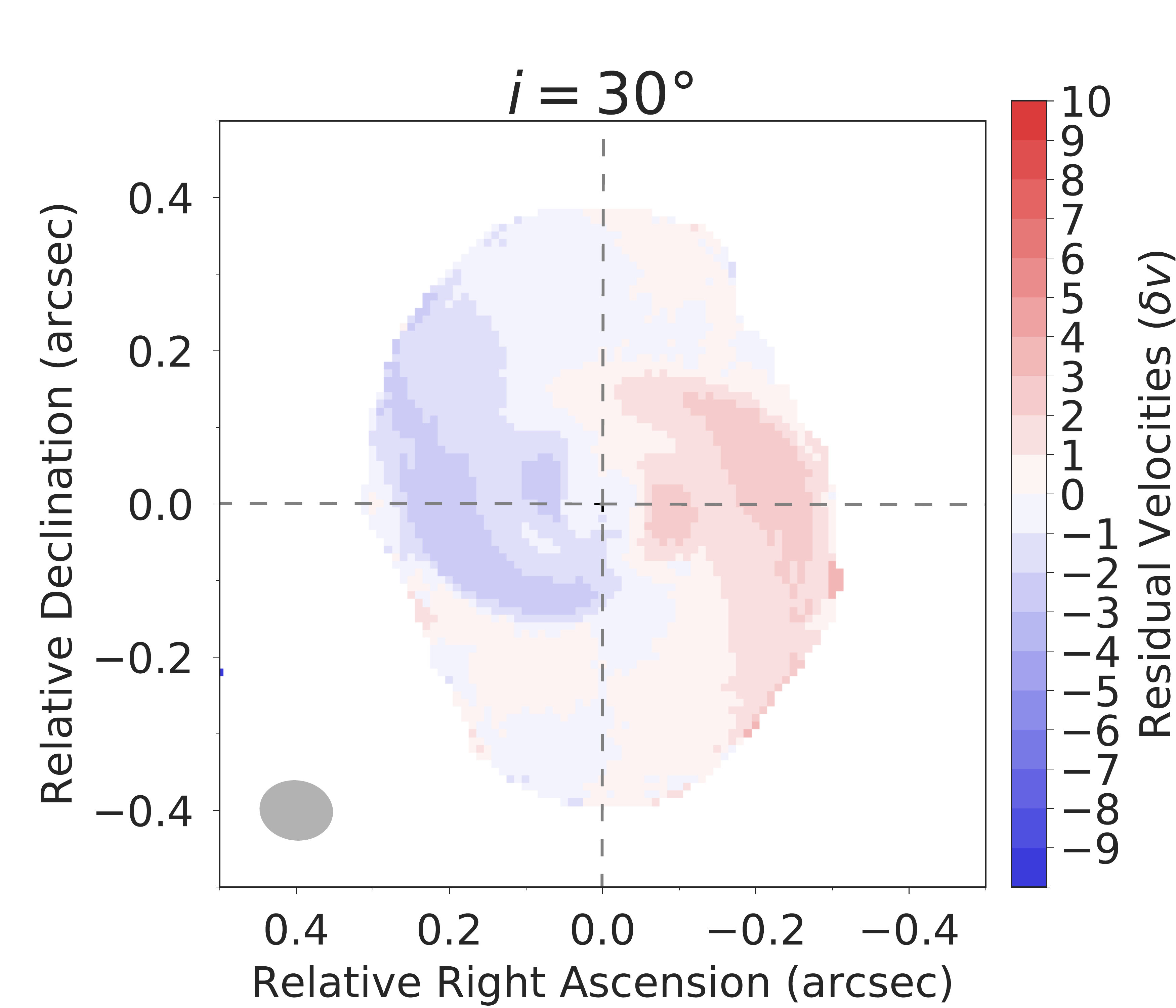}
    \includegraphics[scale=0.18,trim={6cm 0 5.5cm 0},clip]{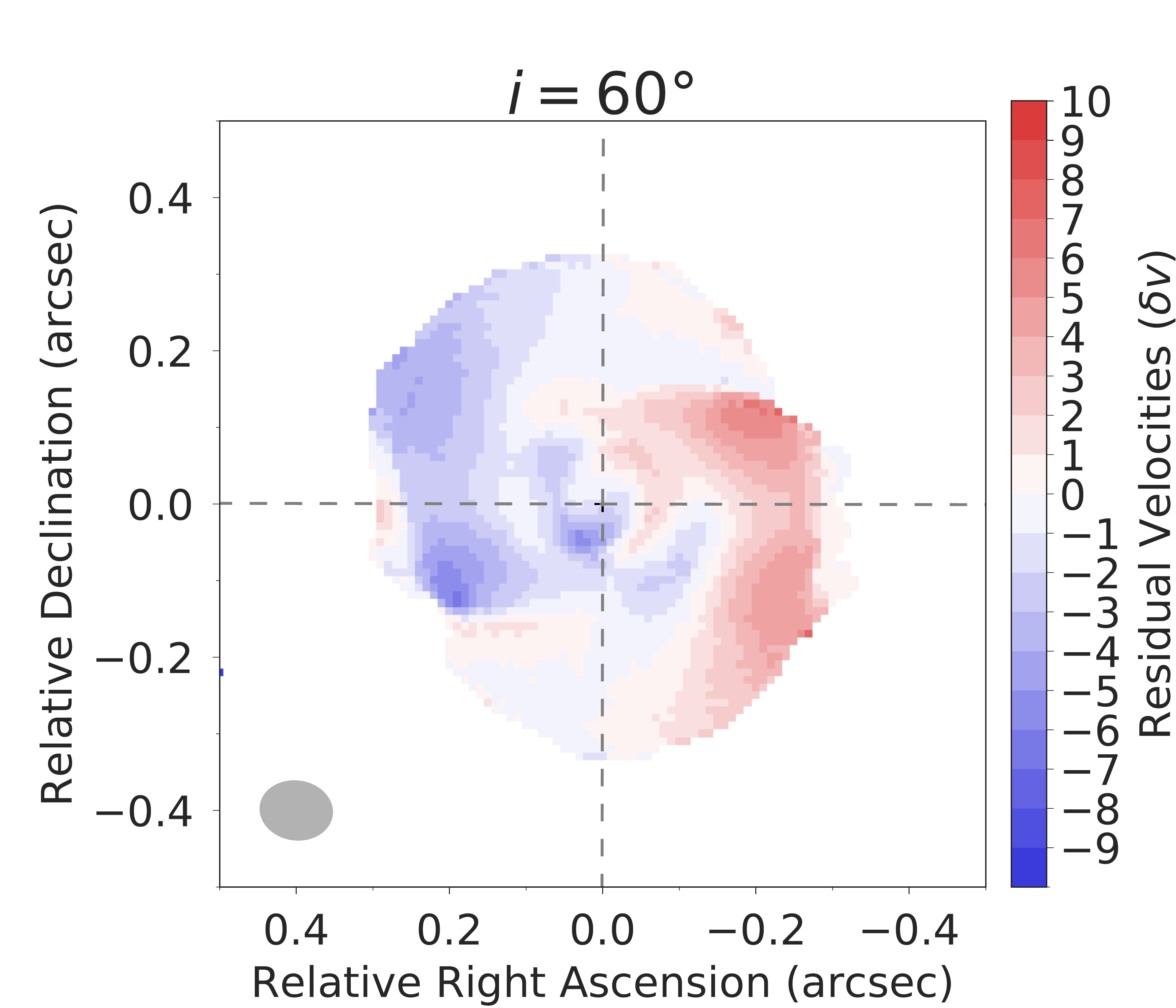}
    \includegraphics[scale=0.18,trim={6cm 0 0 0},clip]{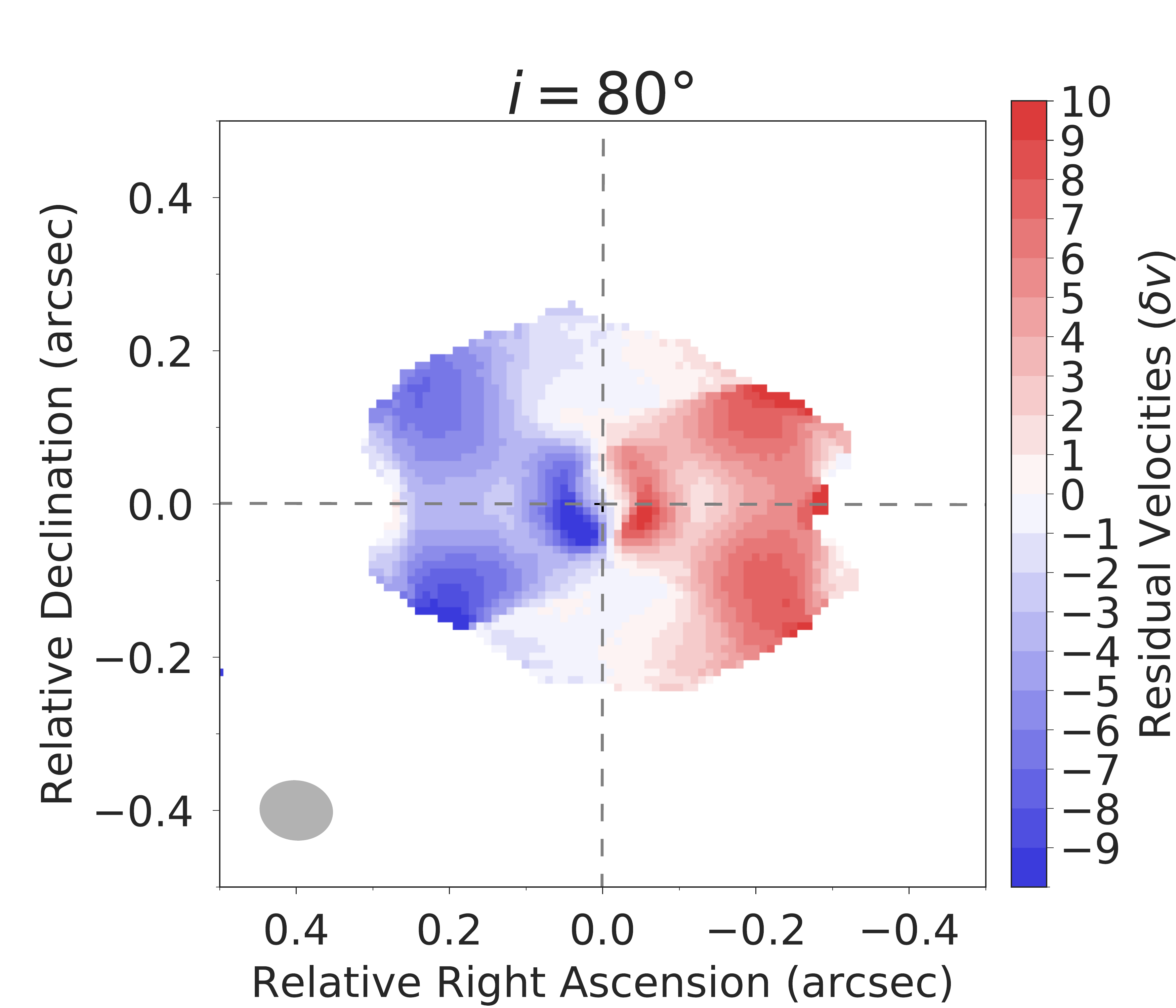}
    \includegraphics[scale=0.18,trim={0 0 5.5cm 0},clip]{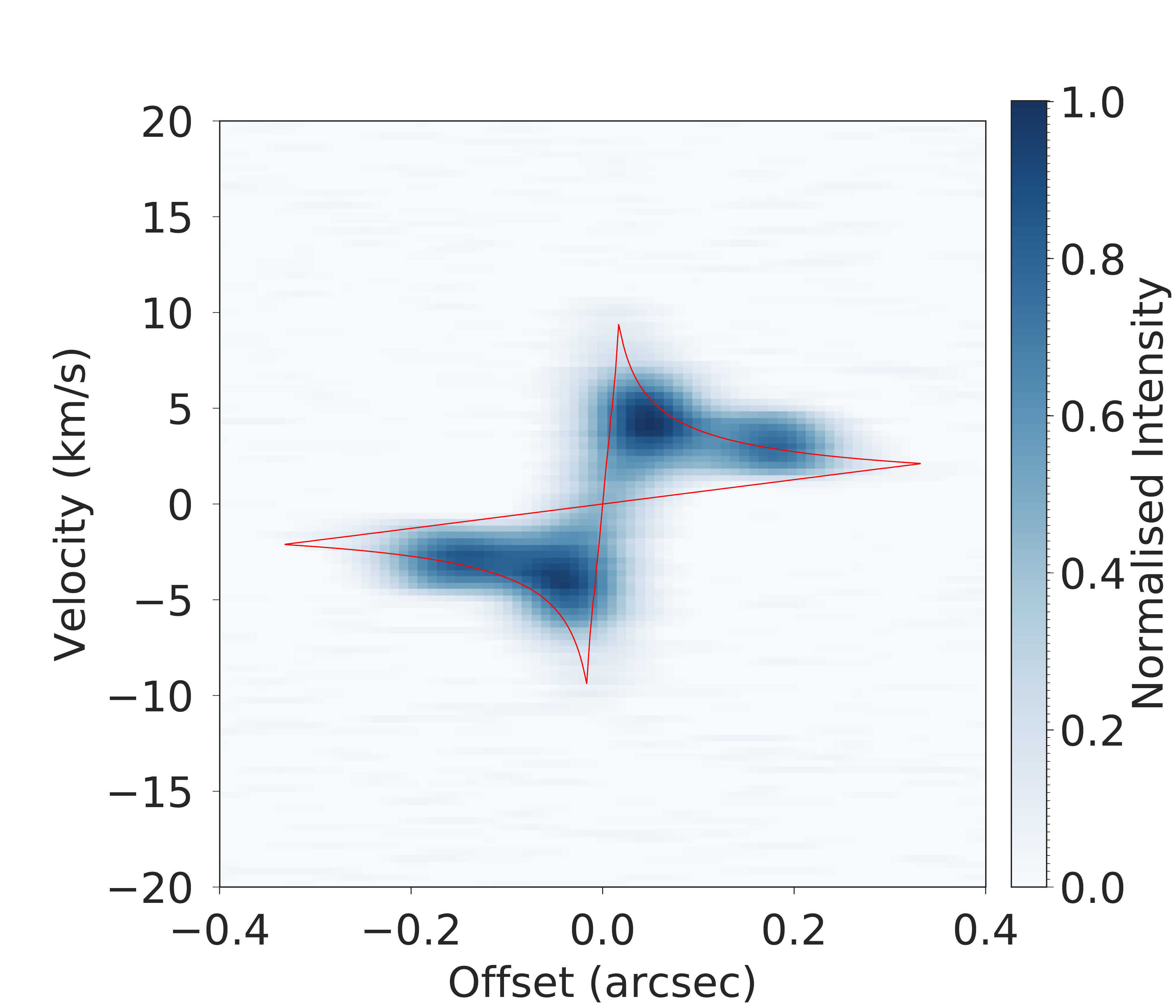}
    \includegraphics[scale=0.18,trim={6cm 0 5.5cm 0},clip]{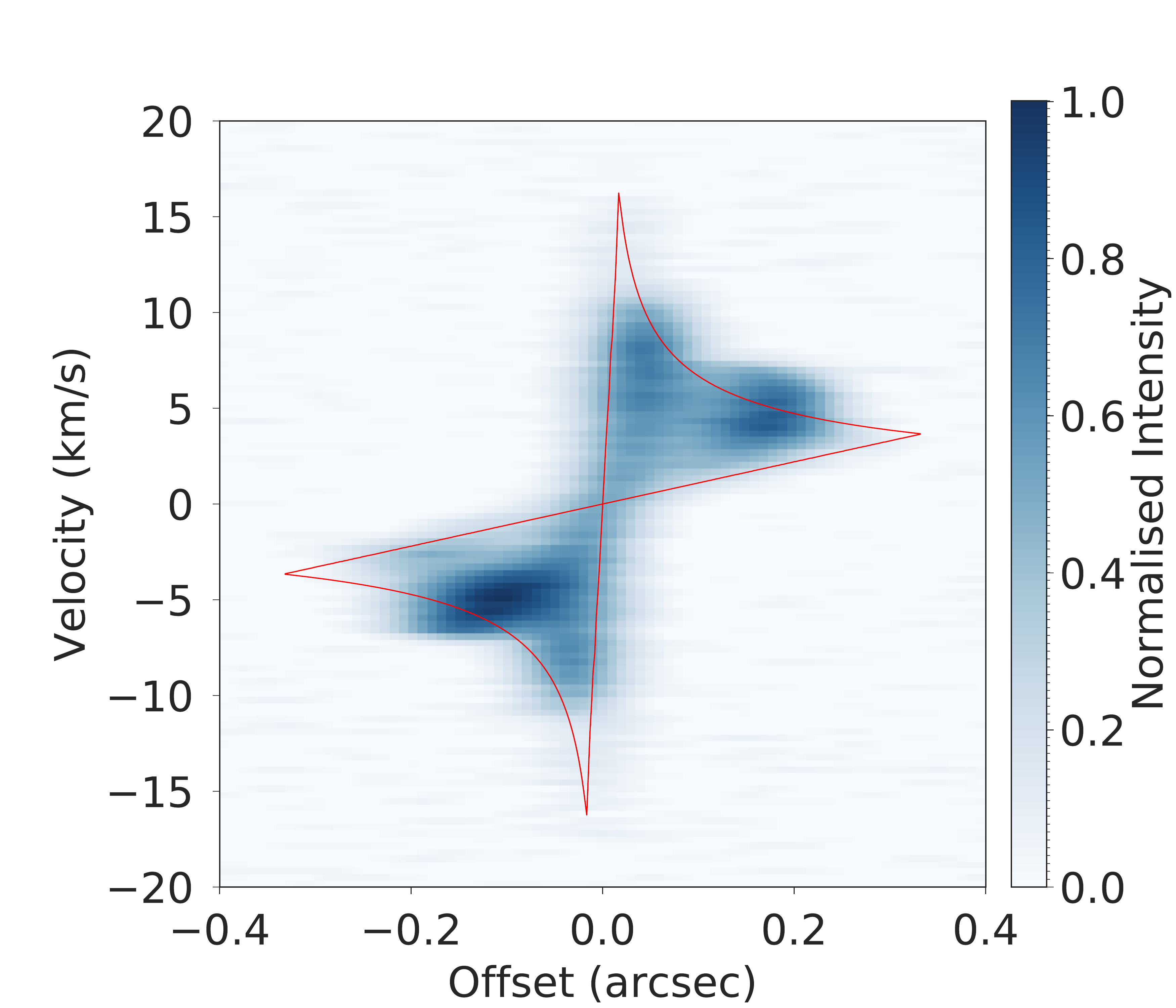}
    \includegraphics[scale=0.18,trim={6cm 0 0 0},clip]{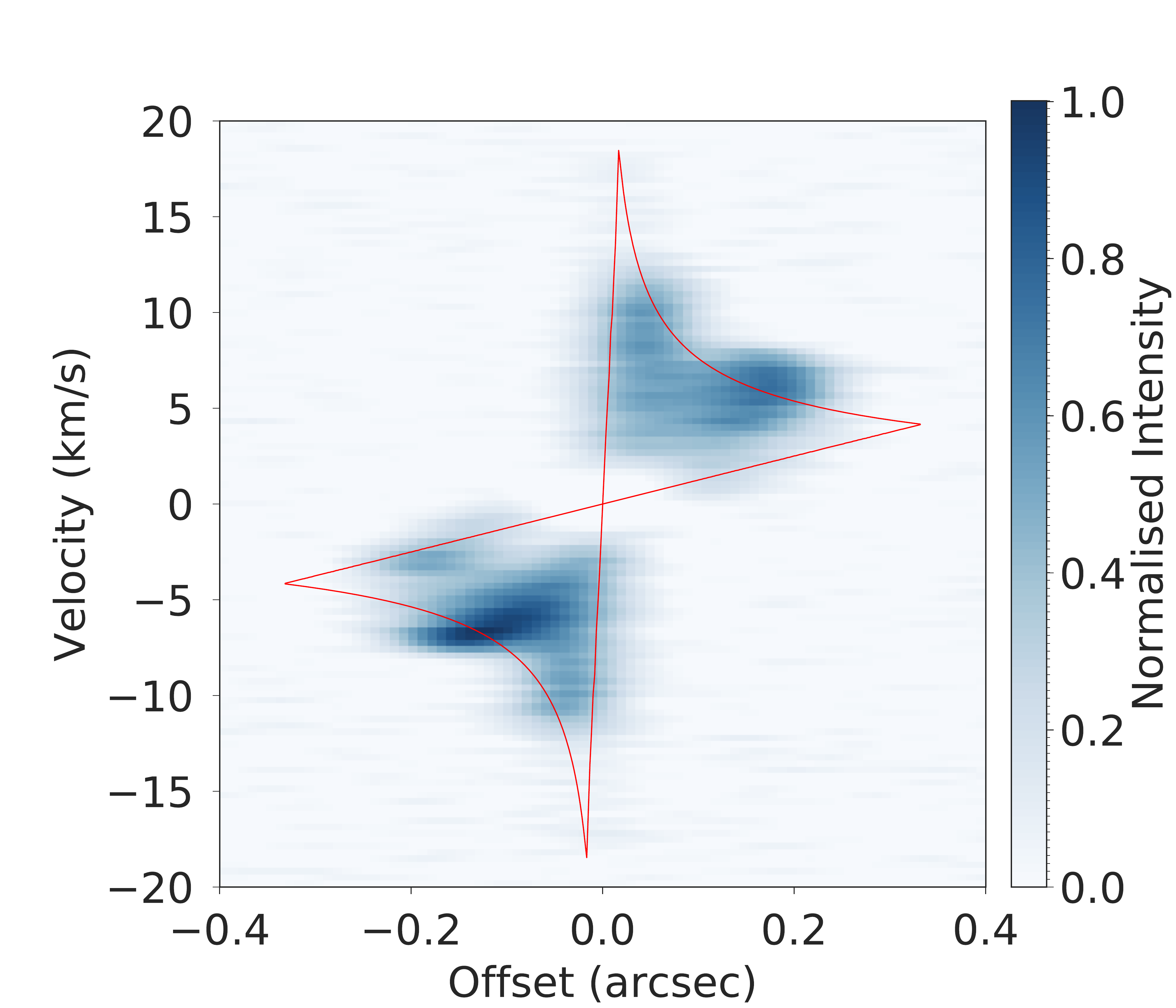}
    \caption{Sensitivity of spectral detection of spirals to disc inclination. The top panels are moment 1 map residuals of the kinematic filtering (given in the units of the spectral resolution $\delta v=0.4$\,km\,s$^{-1}$), and the bottom panels are PV diagrams of our fiducial disc model. Intensity is normalised with respect to the peak value in each diagram. Red lines indicate the expected position-velocity signature of a smooth, Keplerian disc with the same parameters as the model.  For both top and bottom panels: left to right disc inclination is $30^\circ$, $60^\circ$, $80^\circ$.}
    \label{fig:spectralVariations}
\end{figure*}

The usefulness of spectral diagnostics in probing the kinematic perturbation caused by the self-gravity-induced spiral substructure depends on how the spectral resolution of the observation relates to the line-of-sight-projected velocities, and is thus sensitive to disc inclination: an edge-on disc will be easier to resolve spectrally than a face-on disc (if otherwise identical). We thus expect that deviations from Keplerian motion will be much easier to detect in PV diagrams for more edge-on discs. Indeed, this is seen in the bottom panels of Fig. \ref{fig:spectralVariations}, which show PV diagrams of our fiducial disc model inclined at $30^\circ$, $60^\circ$ and $80^\circ$. In each case the red line denotes the Keplerian profile. The signature of spirals is a twin-lobe feature that cuts diagonally across the profile, which if sufficiently resolved should appear as a figure-of-8 like structure, similar to that suggested by \cite{2013MNRAS.433.2064D}. The signature becomes more prominent with increasing inclination.

\smallskip

\begin{figure*}
    \centering
    \includegraphics[scale=0.18,trim={0 0 5.5cm 0},clip]{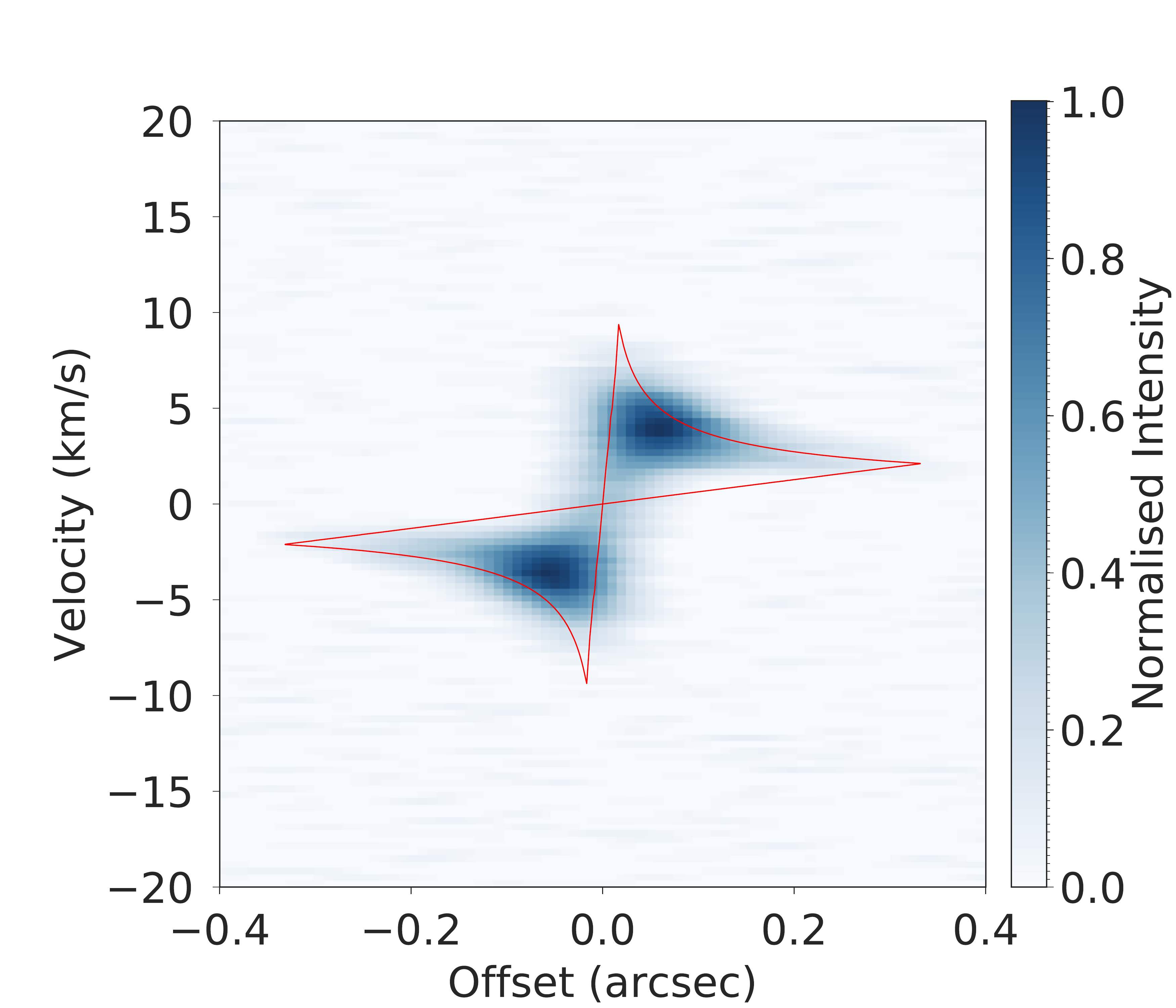}
    \includegraphics[scale=0.18,trim={6cm 0 5.5cm 0},clip]{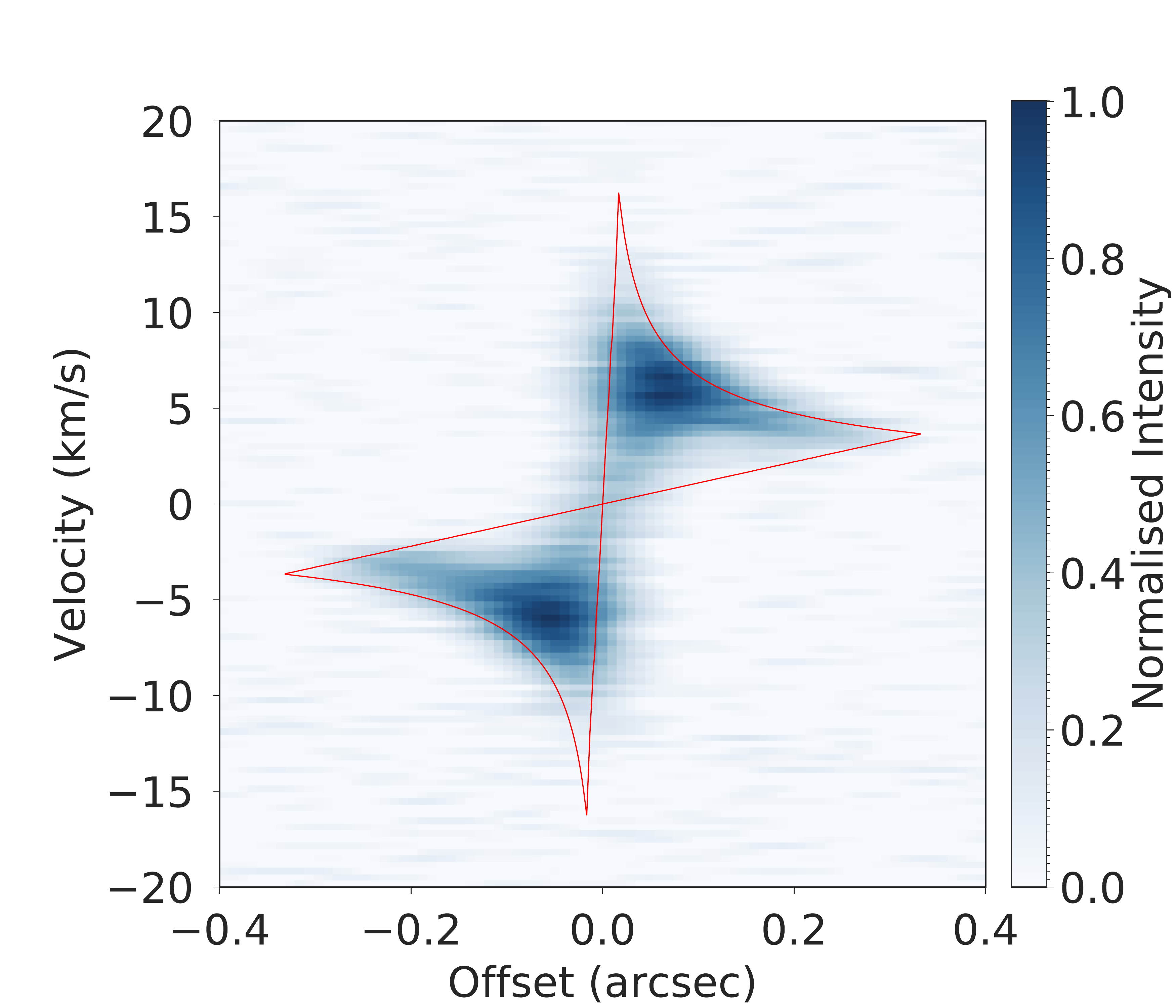}
    \includegraphics[scale=0.18,trim={6cm 0 0 0},clip]{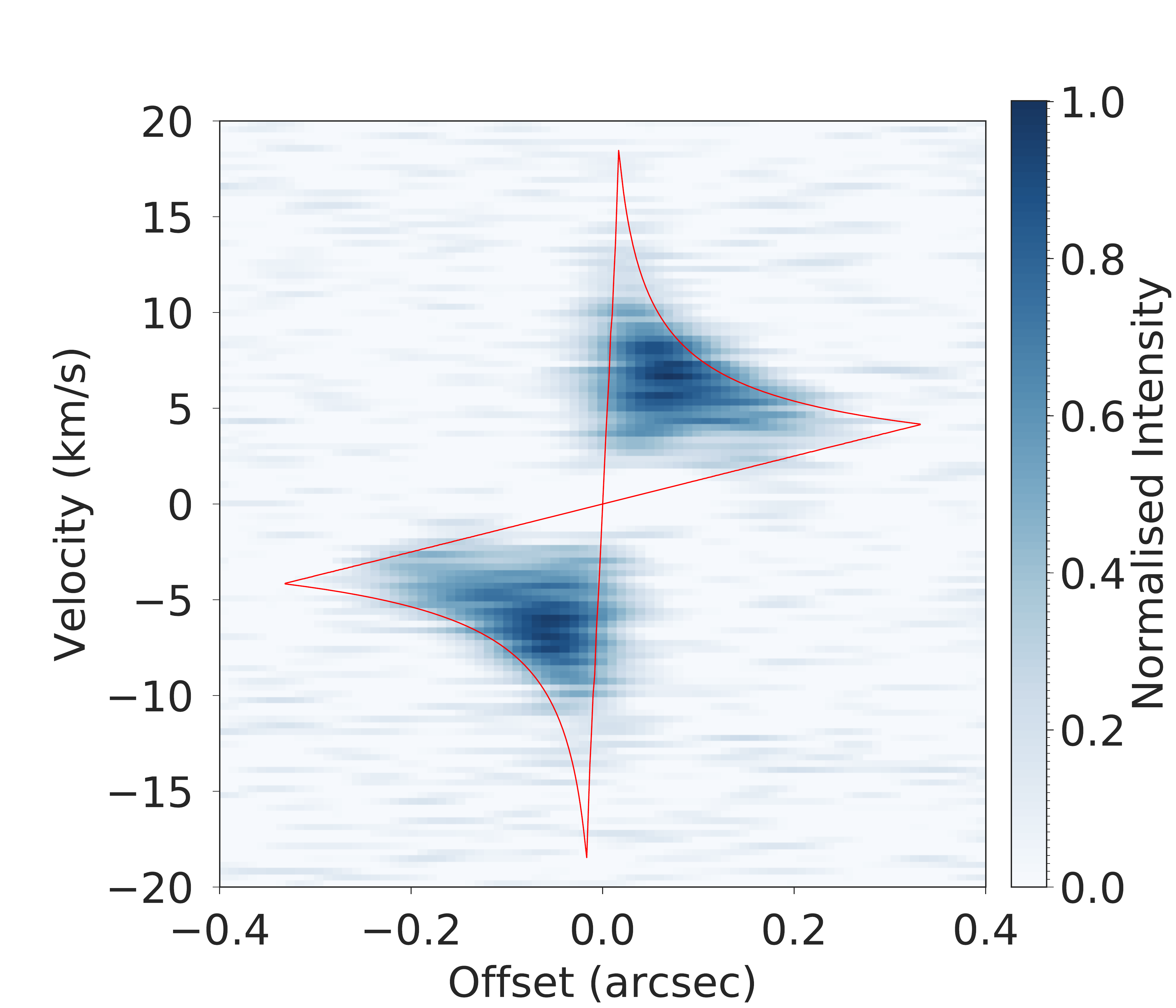}
    \caption{PV diagrams of an axisymmetric disc model, identical to our fiducial disc model except for the lack of spirals. Intensity is normalised with respect to the peak value in each diagram. Left to right disc inclination is $30^\circ$, $60^\circ$, $80^\circ$. Red lines indicate the expected position-velocity signature of a smooth, Keplerian disc with the same parameters as the model. Deviations from this Keplerian profile are discussed in section \ref{sec:spectrally_resolving}.}
    \label{fig:axisymmetricPV}
\end{figure*}

In addition to the spiral signatures, there are other deviations from the Keplerian profile in the PV diagrams, i.e. deviations from the curved red lines. These deviations are seen in the PV diagrams of both spiral (Fig. \ref{fig:spectralVariations}) and axisymmetric disc models (Fig. \ref{fig:axisymmetricPV}). For example, in discs inclined at $30^\circ$ and $60^\circ$, in lower left and middle panels in Fig. \ref{fig:spectralVariations} (and Fig. \ref{fig:axisymmetricPV}), there is emission at low velocities from the inner disc. This is simply due to convolution of the emission with the beam in each velocity channel, that is, an observational effect due to the disc not being perfectly resolved. Noticeably, there is no such emission in the disc inclined at $80^\circ$ (the bottom right panels). This, on the other hand, is an optical depth effect. PV diagrams are produced based on cuts along the major axis of the disc image (i.e., based on the disc position angle; we use the input value from our models). In a fairly face-on disc the cut is made across the ``top'' disc surface, and in a sufficiently inclined disc the cut is along the outermost flared side of the disc. In the disc inclined at $80^\circ$ the line of sight towards the inner disc cuts through a large column of gas with near-zero line-of-sight velocities. Consequently, the near-zero line-of-sight velocity emission from the inner disc region is obscured by the line self-absorption. We discuss the more general effects of optical depth in section \ref{sec:optical_depth}.

\smallskip

Furthermore, there is also missing emission at Keplerian velocities from the outermost disc in the PV diagram of the disc inclined at $80^\circ$, unlike in the less inclined discs. The emission is missing because the PV cut probes the cold disc mid-plane of the outer disc in this case, where the molecule is frozen out. %What is (not) being seen in this case is the cold disc mid-plane in which the molecule is frozen out. 
For a discussion of the observational effects of the molecule freeze out, see section \ref{sec:thermochemical_structure}.

\smallskip

Fig. \ref{fig:spectralVariations} also illustrates how the dependence on inclination is two-fold for the moment 1 maps, as they are also sensitive to the spatial resolution (i.e. beam size). As discussed above, spirals are easier to detect spatially in more face-on discs. Interestingly, the moment 1 maps are also affected by the molecule freeze out. The disc inclined at $30^\circ$ (top left panel in Fig. \ref{fig:spectralVariations}) has an apparent aspect ratio inconsistent with its inclination (and rotation profile). This is due to the molecule freeze out in the outer disc outside of spiral arms. Furthermore, the moment 1 map of the disc inclined at $80^\circ$ (top right panel in Fig. \ref{fig:spectralVariations}) also reveals the absence of the molecule from the disc mid-plane: the radial extent of emission is smaller along the horizontal dashed line (that cuts the disc mid-plane in the outermost disc) than below and above this line (where emission from the hot disc atmosphere is seen).

\smallskip

We conclude from our results that around massive YSOs at $\sim$\,$3$\,kpc spirals will be readily resolved spatially in low-inclined discs ($\lesssim50^\circ$), and readily detected spectrally in PV diagrams in more inclined discs ($\gtrsim60^\circ$). For discs at intermediate inclinations ($\sim$\,$50-60^\circ$), spatial diagnostics aided by moment 1 maps and PV diagrams should unambiguously show spiral substructure. Similar dependencies in detecting substructure on inclination have been found by \cite{2013MNRAS.433.2064D}, albeit for lower mass discs that are much closer than a few kpc.

\smallskip

We note that massive YSOs may be embedded in a natal envelope that may affect observations both by obscuration of emission, and via influencing the kinematic signature of lines (i.e. infall, see \citealt{2016MNRAS.462.4386I}); we do not account for this effect in our models. Kinematic signatures could also be affected by larger-scale filamentary flows in the immediate star-forming environment \citep[see, e.g.,][]{2017MNRAS.467L.120M, 2018MNRAS.478.2505I}.

\smallskip

We find that both spatial and spectral detectability of spiral substructure are only weakly dependent on the rotational offset (rotation angle relative to a fixed disc axis) of spirals in our models (see section \ref{sec:spiralsetup}).

\smallskip

The kinematic perturbation introduced by the spirals in the above models is super-Keplerian, i.e. the spiral pattern is rotating faster than the disc and the matter in spiral overdensities is moving at a velocity that is higher than the local Keplerian velocity. The magnitude of the kinematic perturbation (parameter $\eta$ in our models) is expected to roughly scale with disc-star mass ratio, i.e. with how self-gravitating the disc is \citep{2009MNRAS.393.1157C, forgan_2011}. The above results correspond to a strongly self-gravitating disc ($\eta=0.2$), and we find that it quickly becomes difficult to resolve the perturbation in less self-gravitating discs ($\eta=0.1$). Obviously, this is a function of the spectral resolution of the (synthetic) observation, but the disc physics also plays a role: kinematic perturbation directly determines thermal perturbation in the spirals. The heating due to shocks in spiral arms is expected to be proportional to the kinematic perturbation \citep{2009MNRAS.393.1157C}, as accounted for in our models.

\smallskip

We also explore the possibility of sub-Keplerian kinematic perturbation and find that the twin-lobe feature in the PV diagrams appears similarly as in the super-Keplerian case, but the moment 1 maps become much more difficult to interpret, and spiral signatures are not easily identified. Nevertheless, the pattern speed of spiral density waves in low mass self-gravitating discs is found to be generally super-Keplerian in simulations \citep{2018MNRAS.tmp..327F}. Therefore, spirals should be easy to spectrally resolve in sufficiently inclined and strongly self-gravitating discs.

\subsection{Sensitivity to thermal and chemical structure}
\label{sec:thermochemical_structure}

Molecular abundances can vary greatly within a circumstellar disc, with different molecular species surviving in the gas phase within different regions in the disc. For example, the CH$_3$CN in our fiducial disc model resides in the hot inner disc, the stellar-irradiated warm disc atmosphere and the shock-heated regions of the spiral arms (for the details of the chemical prescription, see section \ref{sec:molecule_prescription}). In the disc mid-plane the molecule is mostly frozen out outside the spirals and outside the snowline at $\sim$\,$300$\,AU.

\smallskip

The molecular abundance in the atmosphere is regulated by the atmospheric heating, a free parameter in our models. In a disc colder than our fiducial one the molecule is frozen out higher into the atmosphere. For sufficiently weaker atmospheric heating, the molecule in the outer disc will practically only trace the spirals, as emission from the low-density upper layers will be too weak. Thus, it could be expected for a cooler disc to make spirals easier to detect as they would have a higher contrast with the rest of the disc. However, we find that varying the atmospheric heating in our fiducial disc model does not affect detection of spirals, since there is already a significant contrast between the shock-heated spiral arms and the rest of the disc. The effect might become important in less massive discs with weaker spiral perturbations, as discussed above.

\subsection{The optical depth}
\label{sec:optical_depth}

The \textsc{torus} radiative transfer calculations give us quick access to the optical depth in both gas and dust in the models. For reference the dust absorption coefficient at frequency $\nu$ in a medium of density $\rho$ is 
\begin{equation}
    \alpha_{\nu}^{\textrm{dust}} = \kappa_{\nu}\rho
\end{equation}
for opacity $\kappa_{\nu}$. Additionally the absorption coefficient in the line is set by the balance between stimulated photoabsorption and induced emission
\begin{equation}
    \alpha_{\nu}^{\textrm{gas}} = \frac{h\nu}{4\pi}\left(n_lB_{lu} - n_uB_{ul}\right)\phi_\nu
    \label{equn:gasAlpha}
\end{equation}
where $n_u$, $n_l$ are the molecule populations in the upper and lower states of the transition, $\phi_\nu$ is the line profile function which is sensitive to both the bulk and microturbulent velocities and $B_{ul}$ and $B_{lu}$ are the Einstein coefficients of stimulated emission and absorption.   We computed the optical depth at the line frequency due to dust and gas (the line self-absorption) from each grid cell on the mid-plane along the trajectory towards the observer in a given model (i.e. the optical depth is inclination dependent).

\smallskip

We first consider the opacity due to dust only in an axisymmetric disc, i.e. a disc without spirals. In this section we allow ourselves to vary the maximum grain size in the distribution to probe the impact of grain growth. We find that there is an inner zone that is optically thick due to dust alone (this extent, i.e. radius, is roughly equivalent for both the K=3 and K=7 J$=13\rightarrow12$ lines, which only differ in frequency by $\sim$\,$0.1$\,GHz). For a maximum grain size $a_{\textrm{max}}$<100\,$\mu$m (minimum grain size 1\,nm, $q=3.3$) this optically thick inner region due to dust alone is around 50\,AU in extent (or less in the case of lower accretion rates, though our inner radius is 50\,AU so we cannot constrain smaller optically thick extents). If $a_{\textrm{max}}$ is increased to around 300--500\,$\mu$m the optically thick region sharply jumps in size to around 100--250\,AU, and remains so for maximum grain sizes up to 1\,cm. These extents are only mildly sensitive to the inclination, at least until the disc is almost edge on and the projected column becomes a lot higher. 

\smallskip

\begin{figure}
    \centering
    \includegraphics[width=7.5cm]{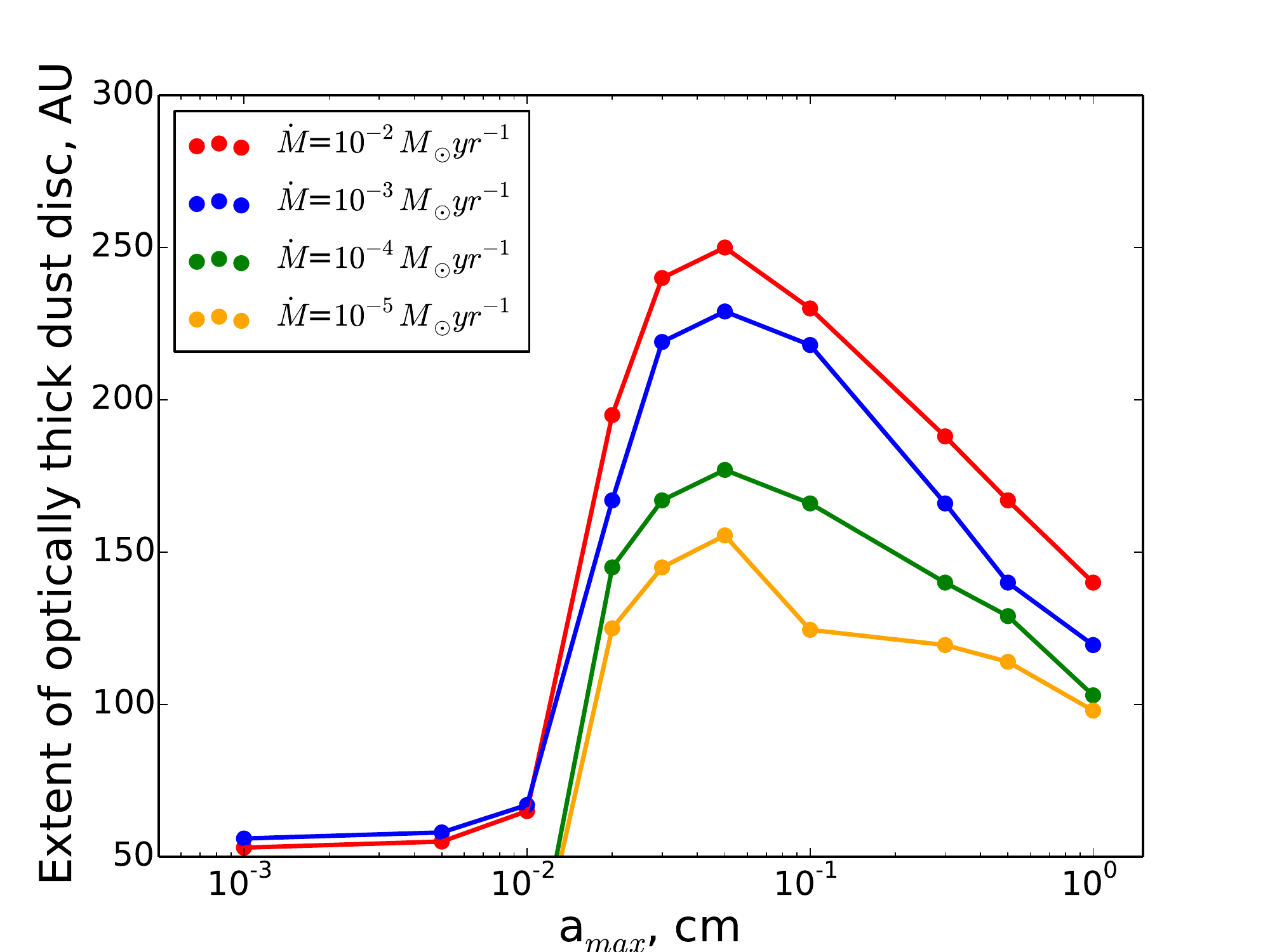}
    \includegraphics[width=7.5cm]{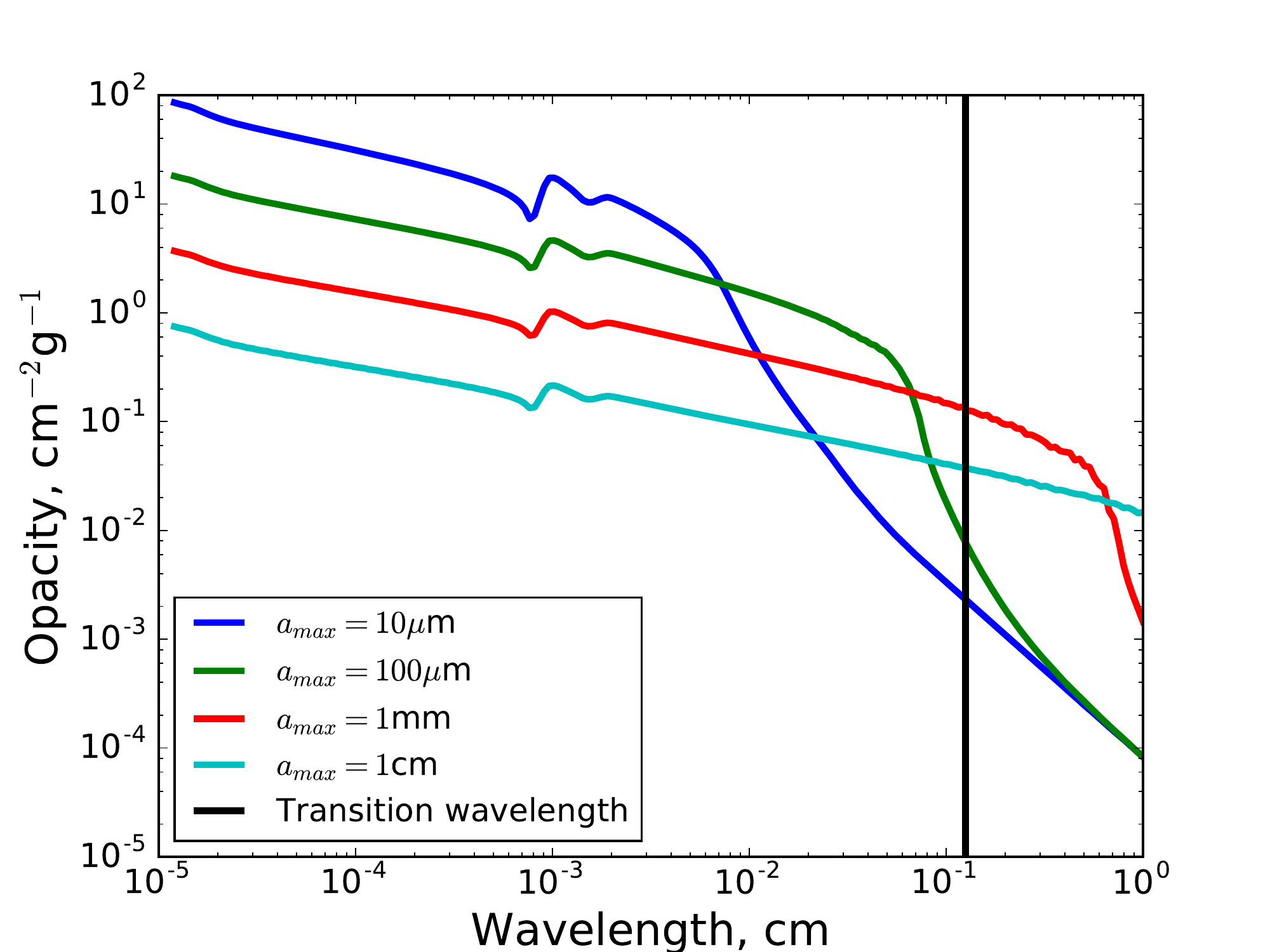}
    \caption{The upper panel shows the extent of axisymmetric regions of discs which are optically thick due to dust alone in the CH$_3$CN K=3, J=13 $\rightarrow$ 12 line. The optical depth at a given radius is computed by moving integrating vertically through the disc. Different lines/colours correspond to different mass accretion rates and hence total disc masses. The lower panel shows the wavelength dependence of the opacity for different $a_{\textrm{max}}$, the behaviour of which at the transition wavelength (black vertical line) determines the form of the upper panel.}
    \label{fig:dustOnlyTau_amax}
\end{figure}

Examples of the extent of the optically thick dust disc as a function of the maximum grain size for a series of axisymmetric discs of different mass accretion rates (and hence total masses) are shown in the upper panel of Figure \ref{fig:dustOnlyTau_amax}. The lower panel of Figure \ref{fig:dustOnlyTau_amax} shows how the opacity varies as a function of wavelength, the behaviour of which at the transition wavelength (black line) explains the form of the upper panel.  Unsurprisingly a larger extent of the disc is optically thick for larger disc masses. %In the highest mass case the zero extent for small $a_{\textrm{max}}$ in the upper panel of Figure \ref{fig:dustOnlyTau_amax} arises because the inner disc is hot enough to sublimate dust. 

\smallskip

We hence expect that discs around massive YSOs observed in these lines will have an optically thick inner region due to dust alone. If this is detected beyond 100\,AU, up to around 250\,AU, it may serve as evidence for widespread grain growth beyond the usual ISM maximum grain size \citep[this is expected in discs, see e.g.][for a review]{2014prpl.conf..339T}.  Note that dense spirals can also induce localised regions that are optically thick in dust at larger radii. The expectation that massive discs around massive YSOs will have such optically thick regions suggests that mass estimates from the continuum are likely to be underestimates \citep[a similar conclusion was reached by][]{2016MNRAS.463..957F}.

\smallskip

The line opacity (equation (\ref{equn:gasAlpha})) can also make the emission optically thick at radii larger than the inner region set by the dust opacity. However, we find that this is difficult to generalise as the line is marginally optically thick/thin. It depends on a plethora of factors including: the chemical and thermal structure of the disc, the viewing angle (and hence the distribution of velocities along the line of sight), the disc mass and so on. The inclination is particularly important to the line opacity, since self-absorption is sensitive to the kinematics along the line of sight. Our fiducial face-on disc is optically thick in the line centre out to around 600--800\,AU when the line opacity is included but at an inclination of 60\,degrees the same model is only optically thick in the line centre out to the extent set by the dust. 

\smallskip

Overall, the K=3, 7 J$=13\rightarrow12$ CH$_3$CN lines cannot be assumed to be optically thick/thin at larger radii in the disc. However, the inner disc is expected to be optically thick due to the dust, with an extent that is sensitive to grain growth above an ISM size distribution.

\subsection{Detecting disc fragmentation}
\label{sec:fragment}
In addition to discs with spiral substructure, we also consider fragmenting discs. Fragments contract into dense and hot clumps of gas and dust that ultimately might become stellar companions or planets.

\smallskip

Given their large distances, observations of massive young stellar objects with ALMA will possess a typical spatial resolution of the order of a few 100s of AU. For a distance of 3\,kpc the particular ALMA configuration we use for our synthetic observations yields spatial resolution of $\sim$\,$270$\,AU. On the other hand, analysis of fragmentation in smoothed particle hydrodynamics simulations of self-gravitating discs around a solar mass star identified fragment radii of a few AU \citep{2017MNRAS.470.2517H}. Even if fragment radii in massive systems are extrapolated to reach 10s of AU, the fragments are not expected to be directly resolved by ALMA. However, they could still be detected appearing as beam size clumps.

\smallskip
	
To explore the possibility of fragment detection, we insert an isothermal clump (see section \ref{sec:fragment_method}) into our fiducial 2-arm spiral disc model with an inclination of 30\,degrees. The clump is placed inside one of the spiral arms, at disc mid-plane, and its velocity is equal to the local bulk velocity. We set the clump radius to 20\,AU. We do not consider smaller clumps because smaller clumps are challenging to resolve in our radiative transfer calculations of the 2000\,AU 3D disc model, even with an adaptive mesh\footnote{The total processing time for a model with a clump is about 5 hours using four cores for the radiative transfer calculations, and a single core for the rest, on a desktop, much longer than the 1-hour processing for models without clumps due to a much higher spatial resolution of the data cubes necessary to resolve the clump.}, and because a 40\,AU-diameter is still only a fraction of the beam, as discussed above. In six disc models with clumps we vary the clump orbital distance (200--800\,AU), total mass (0.1--5\,M$_\odot$) and temperature (150--1400\,K).
	
\smallskip	
	
\begin{figure}
    \centering
    \includegraphics[width=\columnwidth]{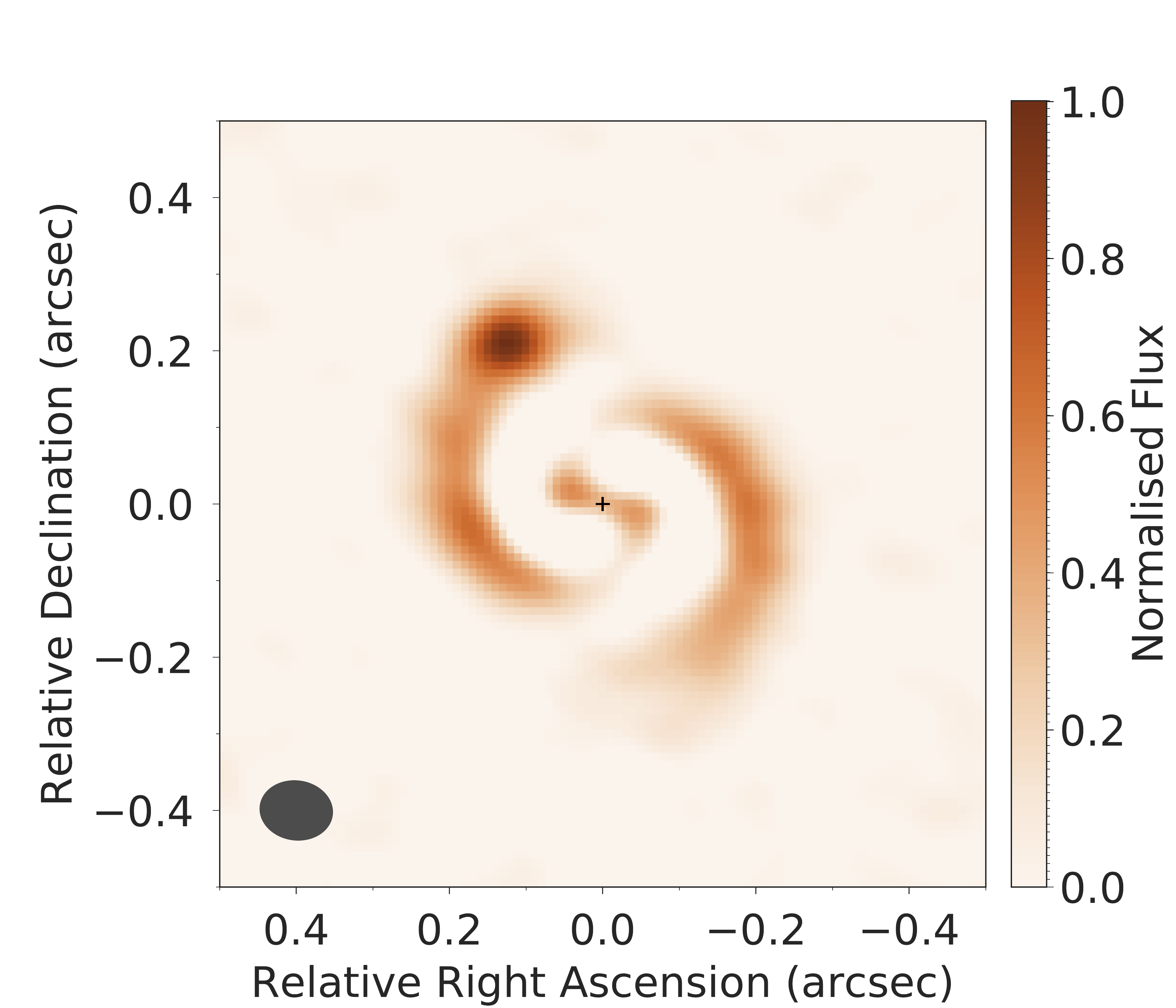}
    \caption{Simulated observed image of our fiducial disc model inclined at 30\,degrees, with a 0.1\,M$_\odot$, 1400\,K fragment inserted at the orbital distance of 800\,AU, in the 239\,GHz continuum. Flux is normalised with respect to the peak value. The relative coordinates of the fragment in the image are (0.12\,arcsec, 0.24\,arcsec). A Gaussian fit has been subtracted from the original image to enhance substructure (see section \ref{sec:spatial_filtering}).}
    \label{fig:fragment}
\end{figure}

Our main findings are that none of the clumps are detectable in the moment 0 nor the moment 1 maps and that the clumps are unambiguously detected in the continuum if they are sufficiently hotter than the surrounding matter, independently of the mass (over the range of masses we consider). Due to shock heating in the spiral arms, even at the orbital distance of 800\,AU the temperature in the spiral arms reaches $\sim$\,$300$\,K at the disc midplane. Consequently, in our synthetic observations the clumps with a temperature of 1400\,K are easily detectable, and those with a temperature of 300\,K are not detectable at all. Again, we find that this is insensitive to the total clump mass in the range of 0.1--5\,M$_\odot$. As an example, we show the continuum image of the model with a 0.1\,M$_\odot$ and 1400\,K clump in Fig. \ref{fig:fragment}.

\smallskip

We therefore conclude that the next generation of ALMA observations towards massive YSOs at $\sim$\,$3$\,kpc distances could expect to detect hotter fragments \citep[i.e. those that have sufficiently contracted,][]{2017MNRAS.470.2517H} in the dust continuum, but not in the moment 0 nor the moment 1 maps of line emission. We note, however, that we have not included kinematics of the matter accreting onto a fragment. We also infer that it would be more likely to detect clumps if they are outside of spiral arms, due to increased contrast. As the spiral arms are expected to be the birthplace of fragments, this refers to the later stages of fragment evolution when the fragment decouples from the spiral density wave to fall onto a Keplerian orbit. Finally, we stress again that caution is required when interpreting substructure in terms of fragmentation, as we have shown in section \ref{sec:spatial_filtering} that image filtering may result in a clumpy structure, which may be difficult to disentangle.

\section{Summary and conclusions}
\label{sec:conclusions}

We have developed a means of quickly generating flexible synthetic observations of self-gravitating discs.  Our approach is not entirely dissimilar to the parametric approach of \cite{2014ApJ...788...59W} that is applied to low mass, axisymmetric discs.  Unlike the time-consuming process of performing synthetic imaging of hydrodynamical simulations, our approach permits us to explore a range of parameters and to understand what influences the detectability of substructure in gravitationally unstable discs around massive YSOs. In our approach we calculate semi-analytic models of such discs, including features such as spirals and clumps. These are then postprocessed with a radiative transfer code to compute synthetic molecular line observations, that are then modified to account for interferometric/observing effects characteristic of ALMA. We draw the following main conclusions from this work. \\

\smallskip

\noindent 1. Spatial and kinematic filtering techniques, such as those described in sections \ref{sec:spatial_filtering} and \ref{sec:kinematic_filtering}, are crucial for detection of substructure in discs around massive YSOs at the kpc distances at which they are observed. \\

\smallskip

\noindent 2. We predict that ALMA will be able to resolve 2-arm spirals at a majority of distances in the 1--5\,kpc range explored here. Substructure in discs with a larger number of spiral arms, i.e. more weakly self-gravitating discs, will be much more difficult to infer. \\

\smallskip

\noindent 3. Spirals at $\sim$\,3\,kpc will be easily resolved spatially in discs inclined up to $\sim$\,50\,degrees. At inclinations of $\sim$\,50--60\,degrees, it will be possible to identify spirals with the help of kinematic diagnostics (moment 1 maps). Above an inclination of 60\,degrees, we find that spirals can be detected only in PV diagrams. \\

\smallskip

\noindent 4. The molecular abundance distribution can in principle be a strong influencing factor for the detection of substructure. It is sensitive to the thermal conditions and molecules may be completely frozen out in the cold dense mid-plane, tracing only the hot disc atmosphere and shock-heated spirals. Weak atmospheric heating may lead to freeze out of the molecule higher into the atmosphere, but we find this not to be important for the detection of spirals, within the parameter space explored here.   \\

\smallskip

\noindent 5. The optical depth of the CH$_3$CN lines considered here varies greatly with the disc inclination and other disc parameters, and the line is often marginally optically thick/thin in the outer disc. We find that the inner disc is always optically thick due to the dust. The extent of the optically thick inner disc may allow the amount of grain growth in the disc to be inferred. \\

\smallskip

\noindent 6. Our modelling suggests that disc fragmentation is unlikely to be detected in the line emission, but that fragments may be detectable in the continuum if they are sufficiently hotter than the surrounding disc material.  Their detectability is not a strong function of their mass, but rather their temperature.   

\smallskip

Our results suggest that upcoming observational campaigns will enable characterisation of the immediate circumstellar enviroments of massive young stars.  Such observations, under certain circumstances, will have the power to begin probing the physical and dynamical conditions of circumstellar discs around MYSOs, and therefore determine the formation mechanisms of massive stars themselves.

\section*{Acknowledgements}

We thank the reviewer for helpful suggestions that improved the manuscript. MRJ is funded by the President's PhD scholarship of the Imperial College London and the "Dositeja" stipend from the Fund for Young Talents of the Serbian Ministry for Youth and Sport. TJH is funded by an Imperial College Junior Research Fellowship. JDI gratefully acknowledges support from the DISCSIM project, grant agreement 341137, funded by the European Research Council under ERC-2013-ADG and support from the STFC (grant number ST/R000549/1).  DF gratefully acknowledges support from the ECOGAL project, grant agreement 291227, funded by the European Research Council under ERC-2011-ADG. CJC acknowledges support from the STFC (grant number ST/M001296/1). CW acknowledges financial support from the University of Leeds. The National Radio Astronomy Observatory is a facility of the National Science Foundation operated under cooperative agreement by Associated Universities, Inc.

\bibliographystyle{mnras}
\bibliography{molecular}

%%%%%%%%%%%%%%%%%%%%%%%%%%%%%%%%%%%%%%%%%%%%%%%%%%
% Don't change these lines
\bsp	% typesetting comment

%\fi
%%%%%%%%%%%%%%%%%%%%%%%%%%%%%%%%%%%%%%%%%%%%%%%%%%

\label{lastpage}
\end{document}